\documentclass{aastex}          
\usepackage{spr-astr-addons}    
\usepackage[hyperindex,breaklinks]{hyperref}

\begin{document}

\title{Regression modeling method of space weather prediction}

\shorttitle{Regression modeling method}
\shortauthors{A.S. Parnowski}

\author{A.S. Parnowski} 
\affil{Space Research Institute of NASU and NSAU\\
prosp. Akad. Glushkova, 40, korp. 4/1, 03680 MSP, Kyiv-187, Ukraine\\
tel: +380933264229, fax: +380445264124\\ e-mail:parnowski@gmail.com}
\email{parnowski@gmail.com}

\begin{abstract}
A regression modeling method of space weather prediction is proposed.
It allows forecasting Dst index up to 6 hours ahead with about $90\%$
correlation. It can also be used for constructing phenomenological models
of interaction between the solar wind and the magnetosphere. With its help
two new geoeffective parameters were found: latitudinal and longitudinal flow
angles of the solar wind. It was shown that Dst index remembers its previous
values for 2000 hours.
\end{abstract}


\section{Introduction}\label{s:Introduction}

The humankind studies space weather for more than 4000 years starting from 
the first mentions of auroras in ancient Chinese literature. The term ``space
weather'' itself exists for almost a century. The official definition adopted
by COSPAR states that ``Space weather describes the physical processes induced
by solar activity that have impact on our terrestrial and space environment, on 
ground based and space technological systems, and on human activities and 
health.'' The first part of this definition actually covers two spatial scales
of space weather, because when we speak about space weather in space, e.g. in
connection with spacecraft failures, we usually mean some local parameters of
the environment, and when we speak about space weather on the Earth, e.g. in 
connection with human health, we usually mean some integral characteristics 
like the geomagnetic indices. Since this article centers on the variations 
of the geomagnetic field, the latter meaning will be used. The second part 
of this definition indicates practical manifestations of space weather. The
impact of the space weather on technological systems is generally accepted
(see \cite{ref:Marubashi}) due to a number of spectacular events like the
superstorm of 1989 when Canada's power grid was disabled for 9 hours and 
numerous spacecraft failures due to ``killer electrons'' causing arcing in 
electronic components, see \cite{ref:Romanova}. The impact on human health,
however, is disputed by most specialists. Nevertheless, the latest reports
(e.g. \cite{ref:KhDim}, \cite{ref:Stoupel}) indicate that there is indeed a
strong correlation between the rate of sudden cardiac deaths and the space
weather.

The space weather problem is twofold. The first aspect is purely practical 
and aims for prediction and, eventually, mitigation of adverse effects of 
space weather. Ideally, this task should be accomplished by launching a vast 
number of spacecraft which will monitor the Sun-Earth region for large-scale 
structures like CMEs. Unfortunately, the resources of the humankind are 
insufficient to produce and maintain such a large space fleet as well as to 
process all the data delivered by these spacecraft. So, today we should use 
the resources at hand, which include a few solar wind spacecraft (ACE, WIND, 
SOHO, and STEREO), magnetospheric spacecraft (CLUSTER, THEMIS), and 
ground-based stations (Intermagnet, MAGDAS, etc.), to develop forecast 
techniques that will be used in future. Thus, we should try to predict space 
weather with what data we have, and we should aim for possibly longer 
prediction times to allow for some kind of countermeasures.

The second aspect is mostly academic and involves study of the processes in
the near-Earth space and, specifically, understanding of interaction between
the solar wind and the magnetosphere. Naturally, improving our knowledge of the 
underlying physics significantly improves predictive capabilities, so
fulfilling the second task will significantly help with the first one. 
Modern conceptions of solar wind-magnetosphere interaction are mostly based 
on phenomenological models constructed in 1960's. However, there are 
numerous problems these models cannot answer. This is largely due to the 
fact that these models were developed at the very beginning of the space era 
when data quality and quantity were immeasurably worse than today. For more 
than 40 years we collected astonishing amounts of data about solar wind 
parameters and geomagnetic activity and now it is time to put them to good 
use.

\section{Possible approaches to space weather prediction}\label{s:Approaches}

Space weather prediction is a challenging and nontrivial activity, see \cite{ref:Li}. The 
most straightforward approach to space weather prediction is studying the 
whole complex chain of physical processes involved in magnetospheric 
dynamics and conjugating them in a global model of the evolution of the 
magnetosphere under the influence of the solar wind. Unfortunately, this is 
not yet possible due to our poor understanding of the physics of the 
interaction between the solar wind and the magnetosphere. For this reason, 
different approaches should be tried.

According to \cite{ref:Kh}, today there are several established methods of
space weather prediction, listed below.

1. Morphological analysis of solar images.

This method provides the longest prediction time (up to a week). Its 
accuracy is unknown since it is used for the academic purposes only. Today 
it is purely manual and thus almost useless for practical implications.

2. Detection of large-scale perturbations in the solar wind, see e.g. \cite{ref:EsFain},
\cite{ref:EFR}.

This method provides a very good prediction time (up to several days), but 
is capable of predicting less than 10\% of the most intense storms.
While it is very inaccurate when used alone, it can prove to be useful
in combination with one of the following short-term methods.

3. Construction of empirical models, see e.g. \cite{ref:BMR}, \cite{ref:Valdivia}, \cite{ref:OM1},
\cite{ref:OM2}, \cite{ref:TL1}, \cite{ref:TL2}, \cite{ref:BG}, \cite{ref:Cid}, \cite{ref:Siscoe}.

This method provides the shortest prediction time (up to 1 hour) with moderate 
accuracy ($\sim 70\%$). Potentially this method could demonstrate far 
better results if the physics behind the magnetic storms was less of a 
mystery.

4. Numerical modeling, see e.g. \cite{ref:Liu}, \cite{ref:MKL}.

This method provides a good prediction time (up to several days) but its
accuracy varies in huge limits. The accuracy of these methods is limited by
their inability to correctly describe plasma instabilities. Besides, the ring 
current can not be described in the framework of ideal MHD, which forms the 
basis of most numerical models. However, they can adequately describe the
motion of e.g. magnetic clouds in the interplanetary environment, but rely on
different methods to detect them.

\begin{figure}[tb]
\includegraphics[width=\columnwidth]{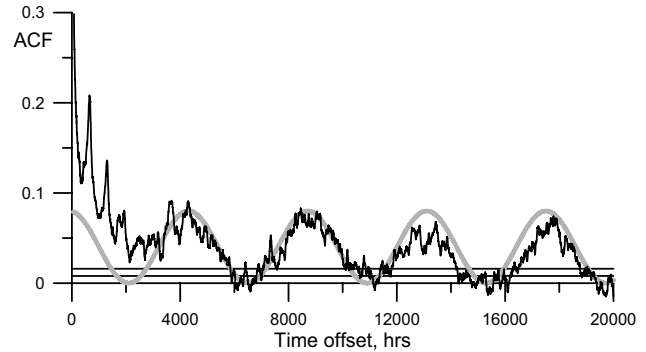}
\caption{Autocorrelation function of Dst. Horizontal lines correspond to top and mean incidental correlation levels in abscence of periodic variations. Gray sine has a period of 1/2 year and depicts seasonal variations}
\label{fig:1}
\end{figure}

\begin{figure}[tb]
\includegraphics[width=\columnwidth]{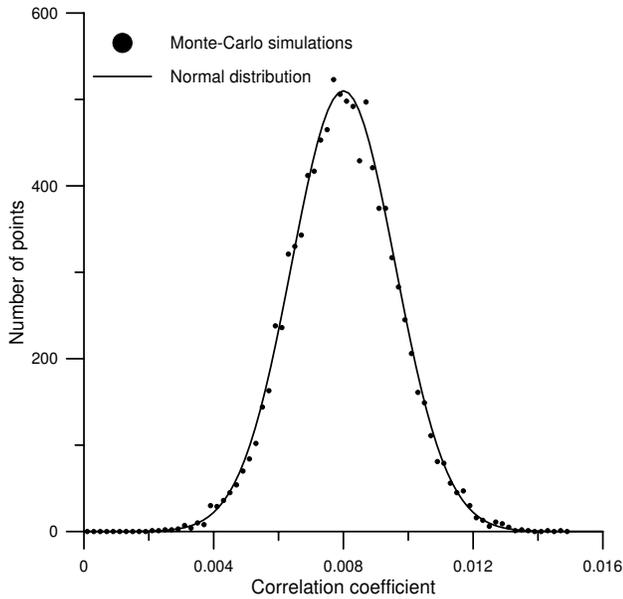}
\caption{Distribution of correlation coefficient of Dst at very large time offsets in abscence of periodic variations}
\label{fig:2}
\end{figure}

\begin{figure}[tb]
\includegraphics[width=\columnwidth]{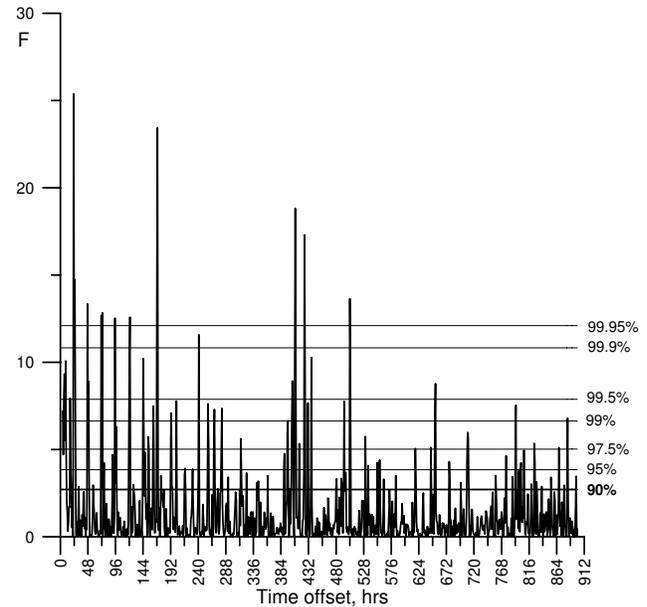}
\caption{Dependence of Fisher significance $F$ of the corresponding term in equation (\ref{eqn:1}) on the time offset for the $1^h$ autocorrelation model}
\label{fig:3}
\end{figure}

5. Multidimensional time series analysis.

This method provides a moderate prediction time (up to several hours) with 
the highest accuracy ($>80\%$). They are very effective and easy to use but 
strongly depend on satellite data availability. These are ``black box'' or 
``input-output'' models, which seek only to reproduce the system's output in 
response to changes of its inputs. The model terms are usually physically 
interpretable and thus useful for construction of new phenomenological 
models. For this reason, this method can not only provide space weather 
forecast per se, but also can improve our knowledge of the physics involved 
and thus increase the efficiency of other methods. 

Further we shall speak about the last method, keeping in mind that its 
results can be used later to assist other methods. First of all, let us 
discuss its existing implementations.

Multidimensional time series analysis can be performed using the methods of 
statistics, signal processing, informatics, fuzzy logic etc. The most widely 
used variations are artificial neural networks (e.g. \cite{ref:Kugblenu}, \cite{ref:Watanabe},
\cite{ref:Wing}, \cite{ref:Pallocchia}), optimization (e.g. \cite{ref:ZhouWei},
\cite{ref:Balikhin}, \cite{ref:HD}), and correlation analysis (e.g. \cite{ref:RB},
\cite{ref:OhYi}, \cite{ref:Wei}, \cite{ref:JW}, \cite{ref:JW5}).
Neural network approach provides short-term predictions up to 4 hours with the
correlation coefficient of 0.79 in the paper by \cite{ref:Wing}. Earlier
implementations of this approach experienced significant difficulties
predicting strong geomagnetic storms with $Kp>5$, but this approach remains one
of the most popular alongside the empirical methods. Optimization approach
seems to be more successful being able to provide 8-hour predictions in the
paper by \cite{ref:HD}. However, in the papers based upon the optimization
methods the volume of the dataset used is insufficient to correctly describe
secular variations of geomagnetic activity. Correlation analysis gives 
interesting results, but it was used solely for developing and constraining 
empirical models (see \cite{ref:JW}). However, most of these methods have a 
common feature: they lead to a regression relationship at some point, so it 
seems natural to skip all the preliminary steps and instantly use the 
regression analysis without unnecessary multiplication of entities. Regression
analysis itself was attempted earlier by \cite{ref:Srivastava}, but it was used
to estimate the probability of intense/super-intense storm occurence depending
on the solar and interplanetary parameters. \cite{ref:Srivastava} was able to
predict 2 of 4 super-intense and 5 of 5 intense CME driven storms during the
1996-2002 period using another 46 CME driven storms to train his model.

Hereafter we propose a new approach, named ``regression modeling'', which 
already allows achieving accurate ($\sim 90\%$) 6 hours ahead forecast of 
the Dst index, which we will use as a quantitative characteristics of space 
weather. This method can be easily extended to predict other geomagnetic
indices like Kp or Ap.

\section{Description of the regression modeling method}\label{s:Description}

The proposed method is statistical, but has some features of empirical 
models. It is based upon the regression analysis and the mathematical 
statistics. In its framework the predicted Dst value is sought in the form
\begin{equation}\label{eqn:1}
D_{st}(j + k) = \sum\limits_{i} {C_i x_i(j)},
\end{equation}
where $j$ is the number of current step (number of hours since Jan 1, 1963), 
$k$ is the prediction length, $C_i$ are the regression coefficients, and 
$x_i$ are the regressors, which are functions and combinations of input 
quantities, which are already measured at the time when prediction is made. 
Values of $C_i$ are determined by least square method (LSM) over a large 
sample of solar wind and geomagnetic data (see next chapter), with equal 
statistical weights of all points. The statistical significance of the 
regressors was determined by Fisher test (F-test) (see \cite{ref:F}, \cite{ref:Hudson}).
This test allows separating significant and insignificant regressors. The 
insignificant parameters are then rejected and the routine is repeated until 
the regression contains only significant regressors. Of course, this method 
does not guarantee that all the significant regressors will enter the 
regression, but physical considerations and brute force in the form of trial
and error provide us with requested reliability. The regressors $x_i$ are generally 
nonlinear, so from the control theory's point of view, this method is able to 
describe discrete dynamical systems with strong nonlinearity. This is an 
essential feature of the regression modeling method.

\begin{figure}[tb]
\includegraphics[width=\columnwidth]{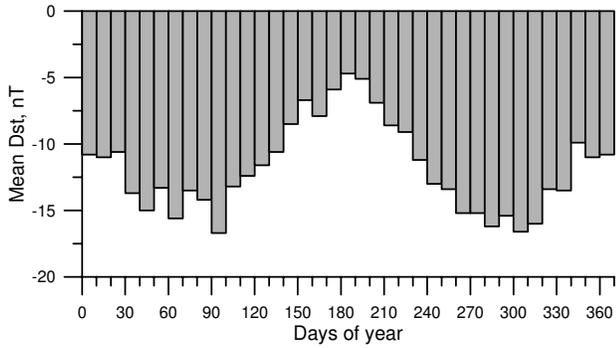}
\caption{Seasonal variation of Dst}
\label{fig:4}
\end{figure}

\begin{figure}[tb]
\includegraphics[width=\columnwidth]{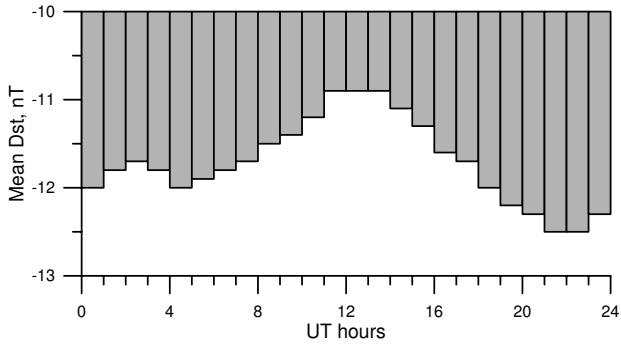}
\caption{Diurnal variation of Dst}
\label{fig:5}
\end{figure}

\begin{figure}[tb]
\includegraphics[width=\columnwidth]{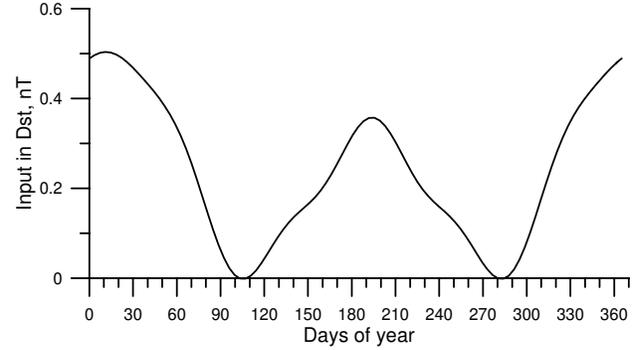}
\caption{Sum of terms directly describing seasonal variation of Dst}
\label{fig:6}
\end{figure}

\begin{figure}[tb]
\includegraphics[width=\columnwidth]{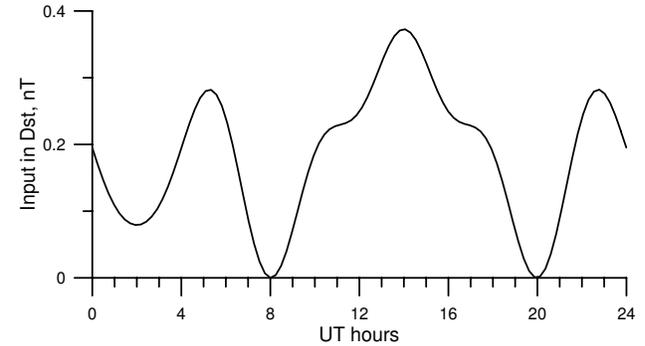}
\caption{Sum of terms directly describing diurnal variation of Dst}
\label{fig:7}
\end{figure}

There is only one manual operation in this method -- selection of regressors 
to be considered. For this purpose all known models, basic physical 
considerations, and random choice are used. Naturally, common sense also 
counts: for example it would be silly to add IMF components in GSE and GSM 
coordinates at the same time. If some regressors $x_i$ appeared to be 
statistically significant, we also checked the significance of products of 
their powers $\prod\limits_{i}{x_i^{p_i}}$, where $p_i$ can be any real number,
including zero, but for practical purposes we used integer values of $p_i$ in
the range from 0 to 6. This yields a very important feature of the regression
analysis: it allows checking the statistical significance of any regressor,
which can be useful for verifying different physical hypotheses. In this sense
we will call a parameter ``geoeffective'' if it appears at least in one
statistically significant regressor.

More details of this method can be found in the article \cite{ref:PUI09}.

\section{Data and routine}\label{s:Routine}

The \cite{ref:OMNI2} database was used. It contains IMF, solar wind 
and geomagnetic data, averaged over 1-hour intervals (49 parameters in 
total, starting from Jan 1, 1963). It was supplemented with provisional Dst 
data, taken from WDC for Geomagnetism (Kyoto). Thus a continuous 44-year Dst 
time series was obtained.

We estimated the geoeffectiveness of a parameter by coefficients and
statistical significances of all regressors, which contain this parameter. This
was done in the following way. After processing the data with the least square 
method, Fisher significance parameter $F$ was determined for each regressor.
All the $F$ values were compared to the values 2.7055, 3.84, 5.02, 6.635,
7.879, 10.83 and 12.1, which correspond to statistical significance of 90, 95,
97.5, 99, 99.5, 99.9 and 99.95\% respectively. Then, insignificant regressors
were rejected and the routine was repeated until all the regressors were
significant. The number of significant regressors depends on the selected
significance threshold. All results given herein correspond to the significance
threshold of 90\%. In contrast to empirical models we do not add fitting
parameters and all the regressors have physical meaning. The described routine
was applied to the sample, obtained from the initial dataset after rejecting
filled values. This sample can be divided into two subsamples, corresponding to
quiet ($Dst>-50 nT$) and perturbed ($Dst\le -50 nT$) conditions.

First, we determined which previous Dst values are statistically significant.
For this purpose, we constructed an autoregression (see details in \cite{ref:EPS08})
\begin{equation}\label{eqn:2}
Dst(j+k)=C_0+\sum\limits_{i=1}^{N}{C_i Dst(j-i+1)},
\end{equation}
where $N$ is the ``age'' of the oldest Dst value. This model alone is not
sufficient to correctly predict Dst, but it sets a basis for the construction
of models that are able to do so. Let us determine the 
maximum reasonable value of $N$. For this purpose, we plot the 
autocorrelation function (ACF) of the Dst index for $k=1$ (see Fig. 
\ref{fig:1}). One can see that ACF tends to a sinusoid with a period close to
half a year. This is caused by seasonal variations. This yields a question: if 
there were no temporal variations, what would ACF tend to at large offsets? 
If the distribution of Dst was normal, the answer would be zero. However, the
distribution is not normal, so ACF can tend to some non-zero quantity. To
determine this quantity we need to remove temporal variations. For this purpose
we need to calculate the ACF of a random sample with the same statistical
characteristics as the Dst sample. The easiest way to get such a sample is to
process the Dst sample with a permutation method, which is widely used for
determination of correlation functions, e.g. in astronomy. This method is based
on random shuffle of the sample. Using this method many times (10000 times in
our case) and calculating the correlation coefficient each time, we get the
distribution of the correlation coefficient by Monte-Carlo method. The
distribution of the correlation coefficient for this sample (Fig. \ref{fig:2})
appeared to be very close to a normal distribution with mean 0.008 and variance
$5.1\cdot 10^{-6}$. The maximum recorded value in 10000 trials was equal to
0.015. The top and the mean values are depicted on Fig. \ref{fig:1} by
horizontal lines. As one can see on Fig. \ref{fig:1}, in reality the correlation
coefficient exceeds this value at most times due to temporal variations. The ACF
crosses the top line for the first time at $\sim 6000$ hours, though the
difference between the ACF and the sine with a half-year period crosses it at
$\sim 2000$ hours, which is about 3.5 27-day periods, so we will assume the
latter value as a rough estimation of $N$. This hints that rather old Dst
values can be quite significant. Besides the half-a-year periodicity one can
also notice the 27-day periodicity, caused by Carrington rotation of the Sun,
which can be taken into account by adding the sunspot number $R$ to the
regression.

\begin{figure}[tb]
\includegraphics[width=\columnwidth]{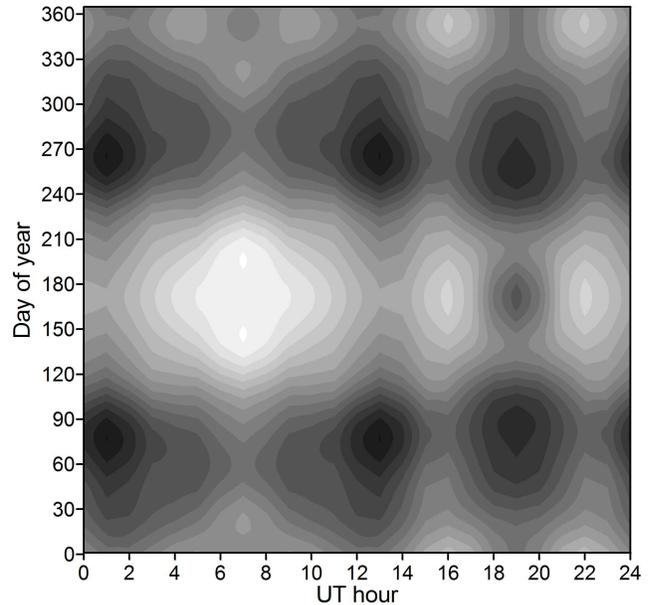}
\caption{Temporal variation of Dst. Darker spots correspond to lower values}
\label{fig:8}
\end{figure}

Let us return to equation (\ref{eqn:2}). Applying the F-test we can determine
which previous Dst values are statistically significant (see Fig. \ref{fig:3}).
We did not search statistically significant Dst values for $N>900$, but it is
possible that there are even older statistically significant values. A similar
situation was reported by \cite{ref:JW} regarding Kp: "the significance is
often quite large for extended periods of time (10-20 days)". As our analysis
shows, Dst index can ``remember'' its previous values for significantly longer 
periods of time. In fact, after adding regressors, corresponding to satellite
data, some of the previous Dst values become insignificant. We found that after
the addition of these regressors there are still statistically significant
values as far as 801 hours ago (33 days and 9 hours) for $k=1$. The statistical
significance of this oldest value is over 99.9\%.

At this point we already have a large number of regressors, describing just the
previous Dst values (autoregression), without satellite data and nonlinear
terms. If we add those, the number of regressors will only increase.

After determining which previous Dst values are statistically significant, we
added all the solar wind parameters available in the OMNI 2 database. Then, we
added nonlinear terms as discussed in Section \ref{s:Description}. After adding
a new regressor, all the significances are recalculated, and some of the old 
regressors can become insignificant. The total number of regressors is about 
150-200. Since it is very large, we will not give here any lists of regressors
or coefficients even for the simplest case $k=1$, but the preliminary list is
given in the paper \cite{ref:KNIT08}.

\section{Identification of new geoeffective parameters}\label{s:Identification}

In this section we will demonstrate how this method can be used for
identification of geoeffective parameters. We will use four parameters as an
example: DOY (day of the year), UT (universal time) and latitudinal and 
longitudinal flow angles of the solar wind.

\begin{figure}[tb]
\includegraphics[width=\columnwidth]{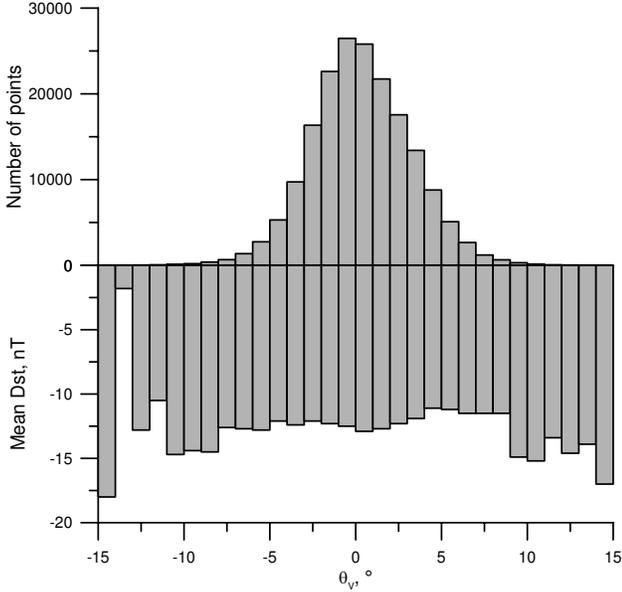}
\caption{Distribution of the latitudinal flow angle and the corresponding mean Dst values}
\label{fig:9}
\end{figure}

\begin{figure}[tb]
\includegraphics[width=\columnwidth]{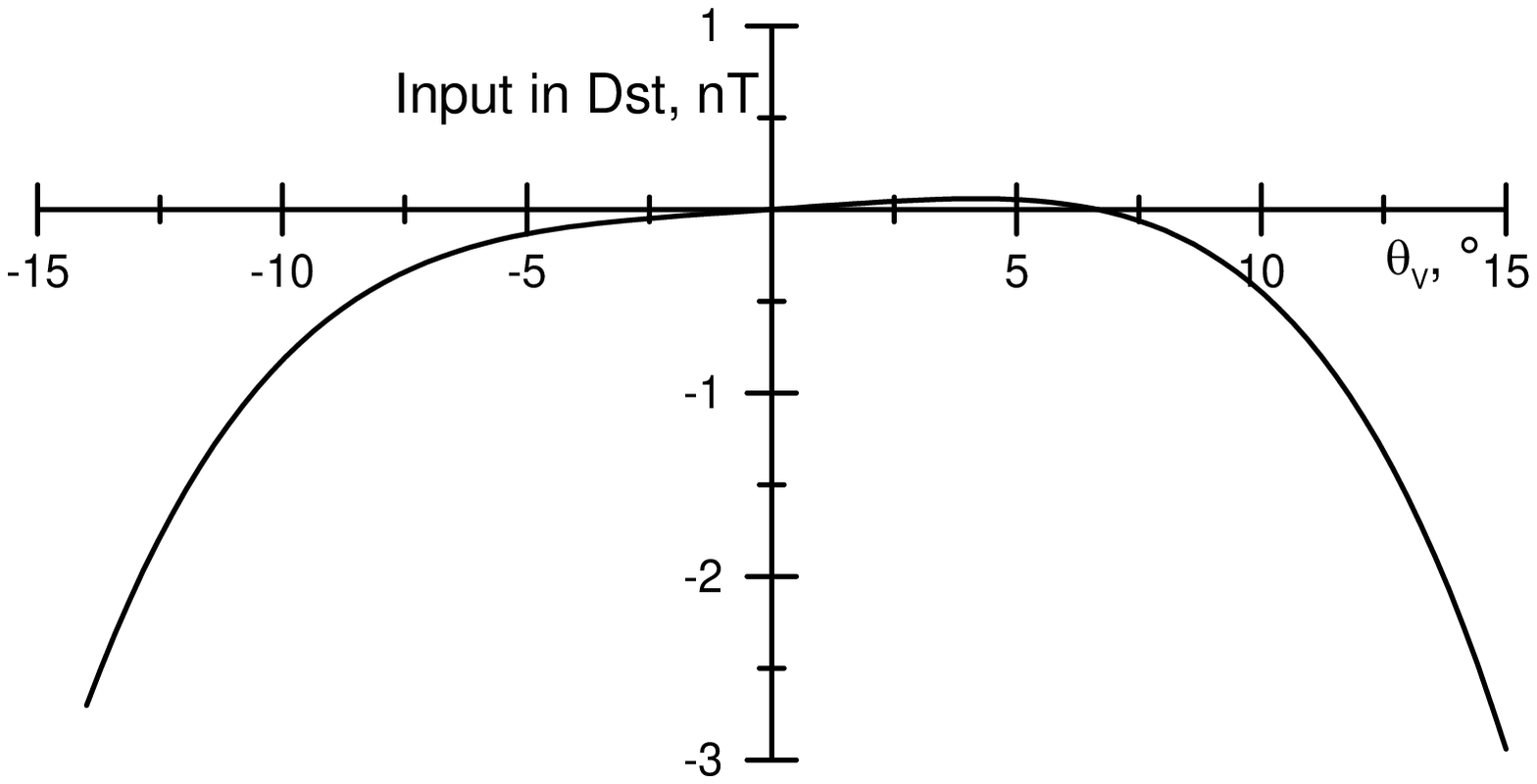}
\caption{Sum of terms describing the latitudinal flow angle}
\label{fig:10}
\end{figure}

\begin{figure}[tb]
\includegraphics[width=\columnwidth]{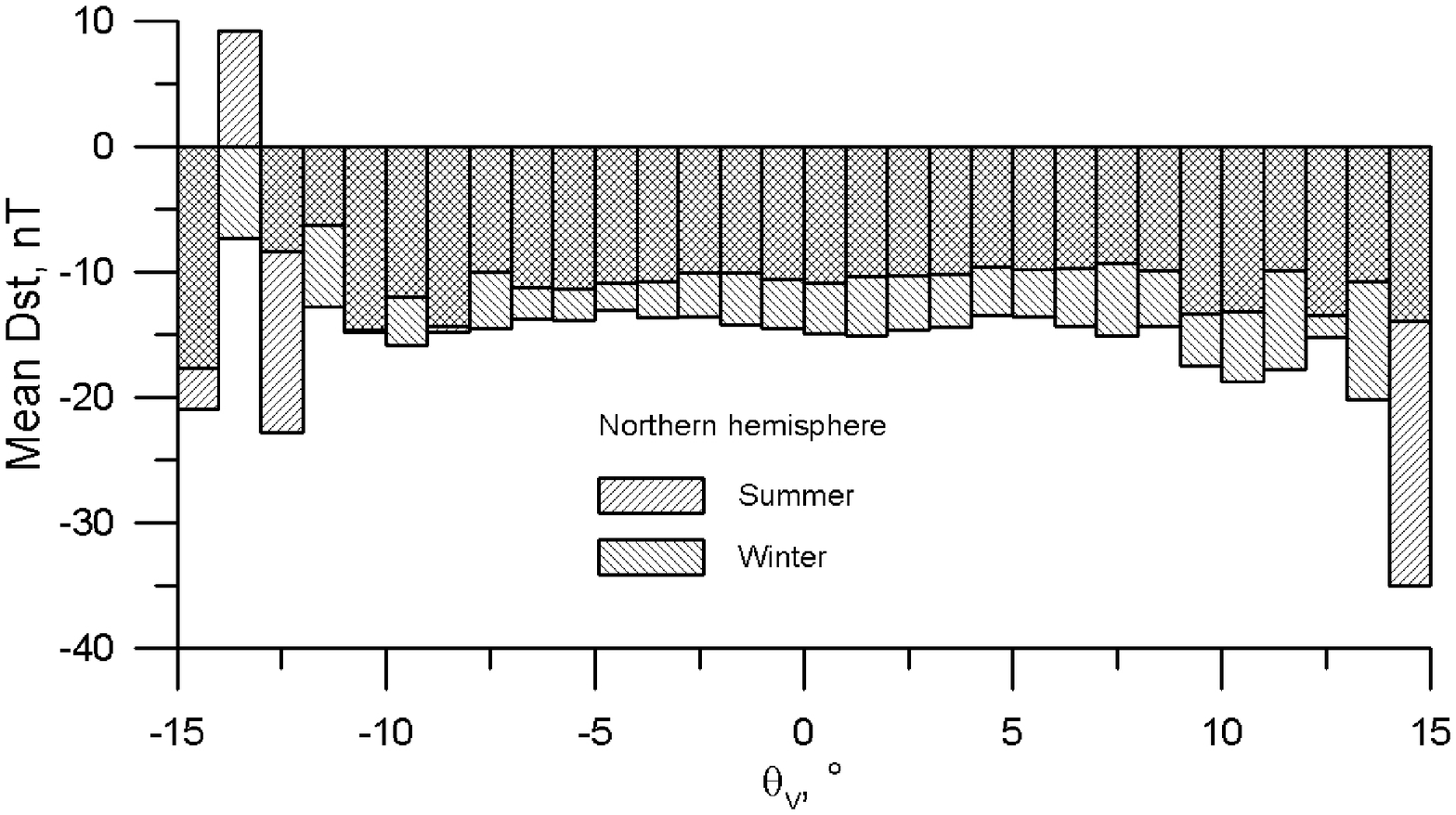}
\caption{Seasonal dependence of latitudinal flow angle's input in Dst}
\label{fig:11}
\end{figure}

On Fig. \ref{fig:1} one could see a clear seasonal dependence of the Dst index.
This dependence was described in many articles, for example by \cite{ref:OM3},
\cite{ref:Lyatsky}, \cite{ref:TM}, and \cite{ref:Cliver}, but the reason behind
it is still disputed. Most authors believe these asymmetries are caused by
either of two cusps turning to the sunlit side due to annual rotation of the
Earth with respect to the Sun. However, \cite{ref:OM3} state that this mechanism
would give only 17\% of observed asymmetry. \cite{ref:TM} connected the diurnal
variations of Dst with an uneven distribution of Dst network stations. Let us
use this known effect to validate our method. 

If we select two subsamples, corresponding to summer and winter in northern 
hemisphere, bounded by vernal and autumnal equinoctia, and verify the
hypothesis that the difference between the corresponding average Dst values 
is statistically significant using a one-sided Student test, we obtain 
$t_\infty=80.264$, which is well over 99.95\% significant. Values of $t_\infty$
corresponding to 99 and 99.95\% significance levels are equal to 2.334 and 3.31
respectively. For diurnal asymmetry Student test gives $t_\infty=8.774$, which
corresponds to significance level of more than 99.95\%. Note that formally
Student test is applicable when Dst is normally distributed. In fact, the
distribution of Dst is slightly asymmetric, but taking into account the
obtained values of $t_\infty$, we can be sure in qualitative conclusions made.
Figs. \ref{fig:4} and \ref{fig:5} show the histograms of seasonal and diurnal
variations of Dst index.

Taking this known geoeffective factor as an example we demonstrate how easily
one can take it into account using regression approach. To do so one should
simply add terms $a_1(j)=\sin((j-1920)\pi/4383)$ and
$a_2(j)=\cos((j-1920)\pi/4383)$ into the regression. Here $j$ is once again the
number of hours since Jan 1, 1963, 1920 is the number of hours between the
beginning of the year and the vernal equinox, and 4383 is the number of hours
in half a year. The first of these terms is significant and describes
summer/winter asymmetry, and the second one (which appears statistically
insignificant) describes an absent spring/autumn asymmetry. Likewise, for
diurnal asymmetry the corresponding terms will be $b_1(j)=\sin((j-2)\pi/12)$
and $b_2(j)=\cos((j-2)\pi/12)$. Here 2 is the time difference between UT and
the northern geomagnetic pole, and 12 is the number of hours in half a day.
Both these terms are significant.

\begin{figure}[tb]
\includegraphics[width=\columnwidth]{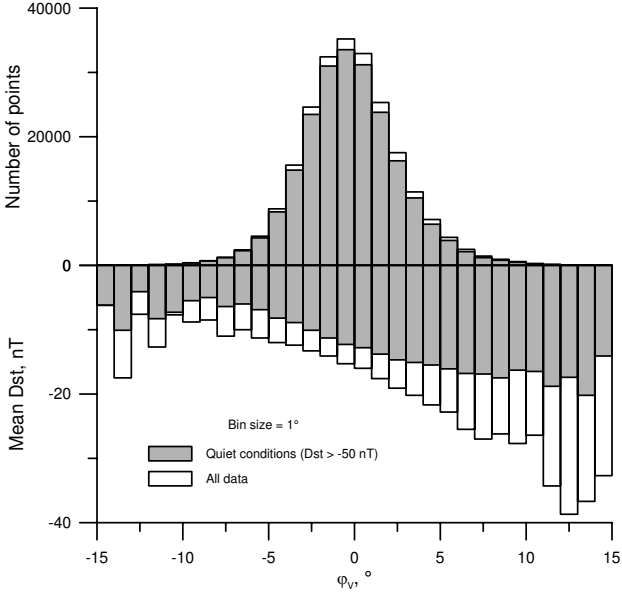}
\caption{Distribution of the longitudinal flow angle and the corresponding mean Dst values. Grey columns correspond to quiet conditions, white columns -- to all data}
\label{fig:12}
\end{figure}

The coefficient of the $a_1(j)$ term is 30 times less than the observed
difference between mean Dst values of summer and winter subsamples. This can be
explained in the following way: there are other regressors, which depend on
parameters with statistically significant summer/winter asymmetry, e.g.
previous Dst values. They provide the lion share of summer/winter asymmetry of
Dst. A good example of such a regressor is the sunspot number $R$, which
describes the 27-day periodicity. Nevertheless, there is a small difference
which can not be expressed with these terms. Including it into regression, we
obtain these statistically significant regressors.
To further illustrate this point, let us consider as an example a value
$X=const+A\sin\omega t$. In the regression it will look like
$X_{n+1}=X_n+A\left[\sin\omega(t+\Delta t) - \sin\omega t\right]
= X_n+A\left[ (\cos\omega\Delta t - 1) \sin\omega t
+ \cos\omega t\sin\omega\Delta t\right]$. The first term in brackets is of
order $(\omega\Delta t)^{2}$, and the second -- $\omega\Delta t$ in the natural
assumption that $\omega\Delta t\ll 1$. So, it will seem that the coefficient is
$A\omega\Delta t$ rather than $A$. Note that this is just an example and has
nothing to do with actual regressors.

\begin{figure}[tb]
\includegraphics[width=\columnwidth]{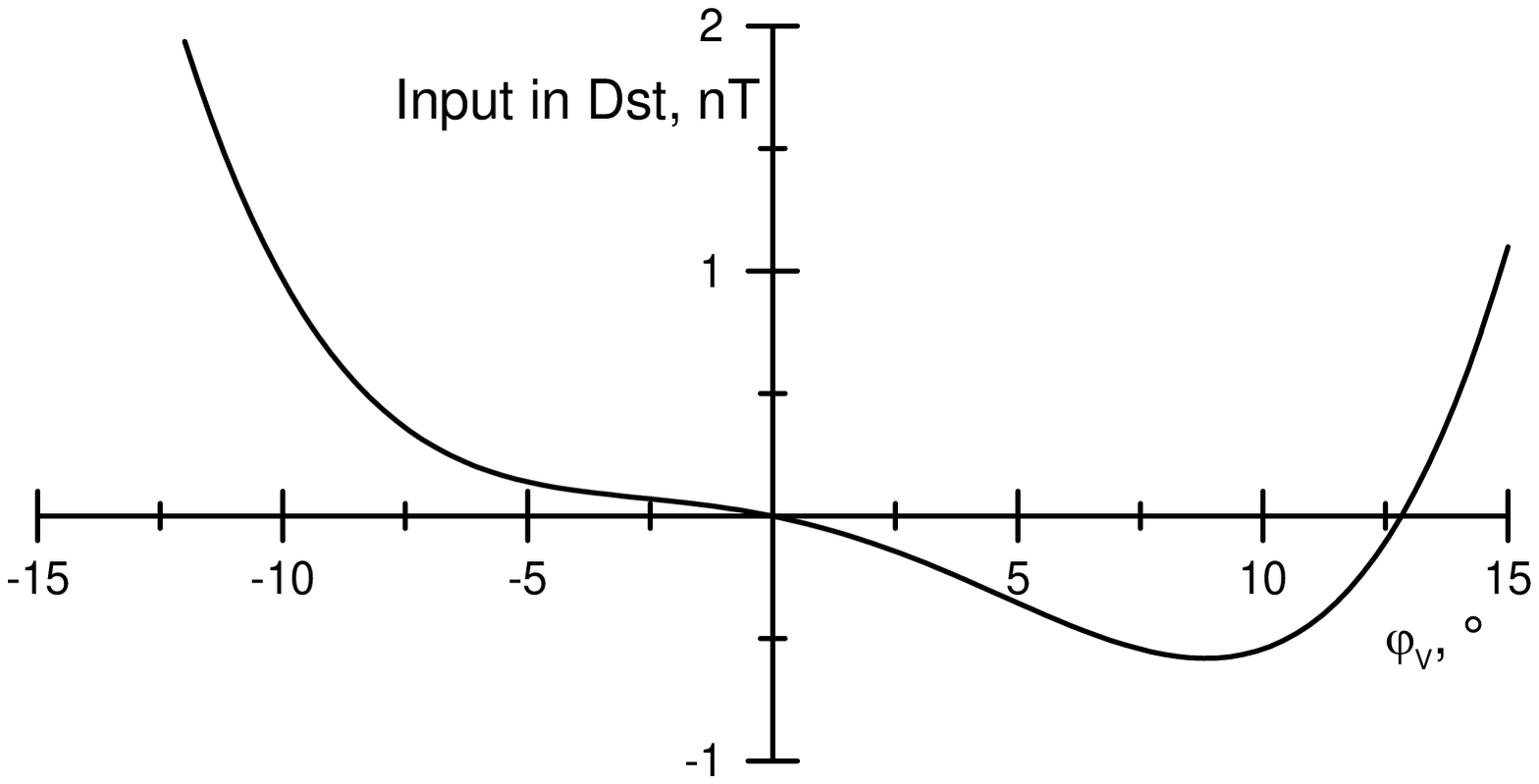}
\caption{Sum of terms describing the longitudinal flow angle}
\label{fig:13}
\end{figure}

\begin{figure}[tb]
\includegraphics[width=\columnwidth]{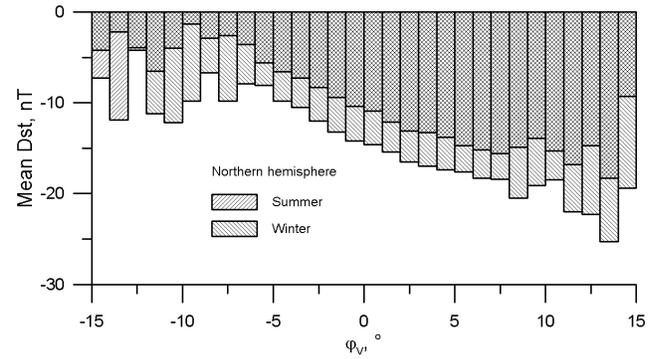}
\caption{Seasonal dependence of longitudinal flow angle's input in Dst}
\label{fig:14}
\end{figure}

However if we look on the distribution of mean Dst values vs. time of the year
(Fig. \ref{fig:4}), we see a much more complicated pattern of seasonal
variations of geomagnetic activity. Among other features there is a strong
asymmetry between summer and winter on one side and spring and autumn on the
other. To take it into account we introduced additional terms into our
regression, which are powers of $a_1(j)$ and their products with powers of 
$a_2(j)$. The sum of regressors with the corresponding coefficients, depicted
on Fig. \ref{fig:6}, is very similar to Fig. \ref{fig:4}. Note that Fig. 
\ref{fig:6} was obtained independently from Fig. \ref{fig:4}.

We did the same thing with the diurnal asymmetry. The distribution is plotted
on Fig. \ref{fig:5}, and the sum of regressors with the corresponding
coefficients -- on Fig. \ref{fig:7}. The term $a_1(j)\cdot b_1(j)$ is also
significant and should be included in the regression. After this we obtained a
joint distribution of semiannual and diurnal variations of Dst index, plotted
on Fig. \ref{fig:8}. It contains 18 regressors. Increasing the number of
regressors describing temporal variations of geomagnetic activity we can
improve the accuracy of this distribution. In particular, one could add 11-year
and 22-year solar cycles, higher powers of $a_i(j)$ and $b_i(j)$ etc.

Thus, we demonstrated how easily one can take into account new geoeffective
parameters in this method's framework.

\begin{figure}[tb]
\includegraphics[width=\columnwidth]{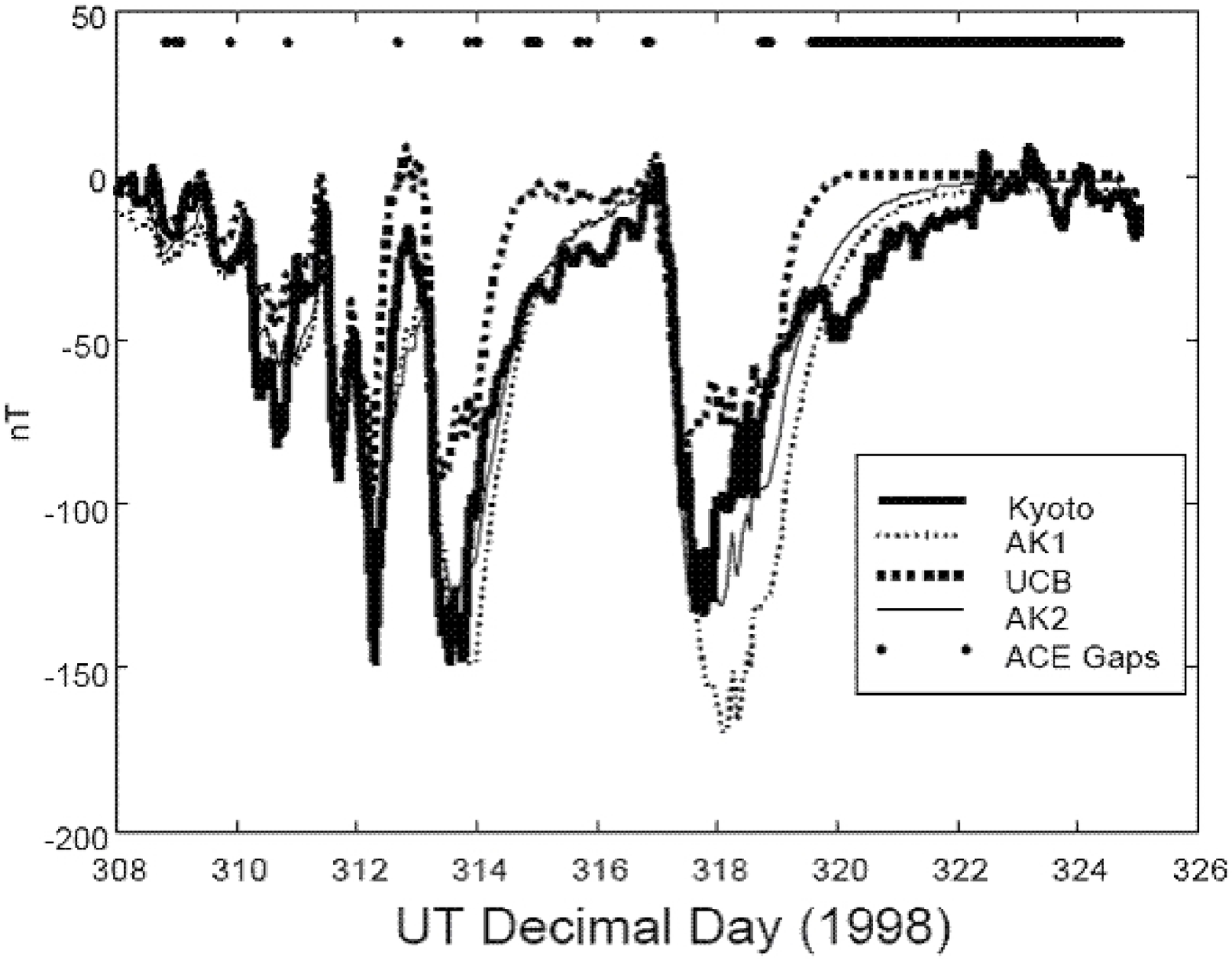}
\includegraphics[width=\columnwidth]{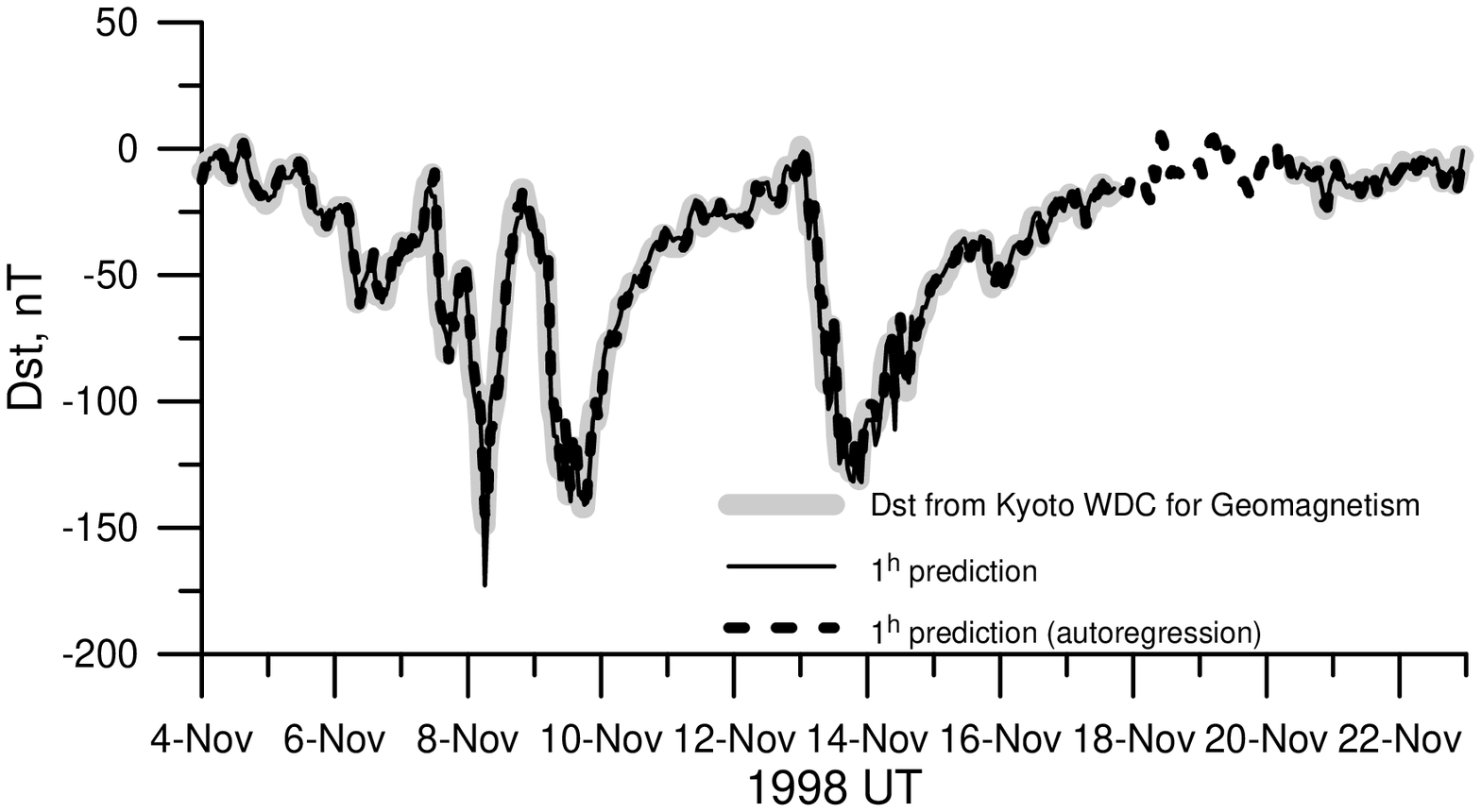}
\caption{Comparison between predictions results of \cite{ref:OM1} (top) and ours (bottom) 1 hour ahead. The following designations are used: 'Kyoto' -- official Dst index, available at Kyoto WDC for Geomagnetism; 'AK1' -- prediction based on the model of \cite{ref:BMR} with re-calculated coefficients; 'UCB' -- prediction based on the model of \cite{ref:FL}; 'AK2' -- prediction based on the model of \cite{ref:OM2}; 'ACE Gaps' refer to the top line, indicating the availability of solar wind data measured by ACE satellite}
\label{fig:15}
\end{figure}

\begin{figure}[tb]
\includegraphics[width=\columnwidth]{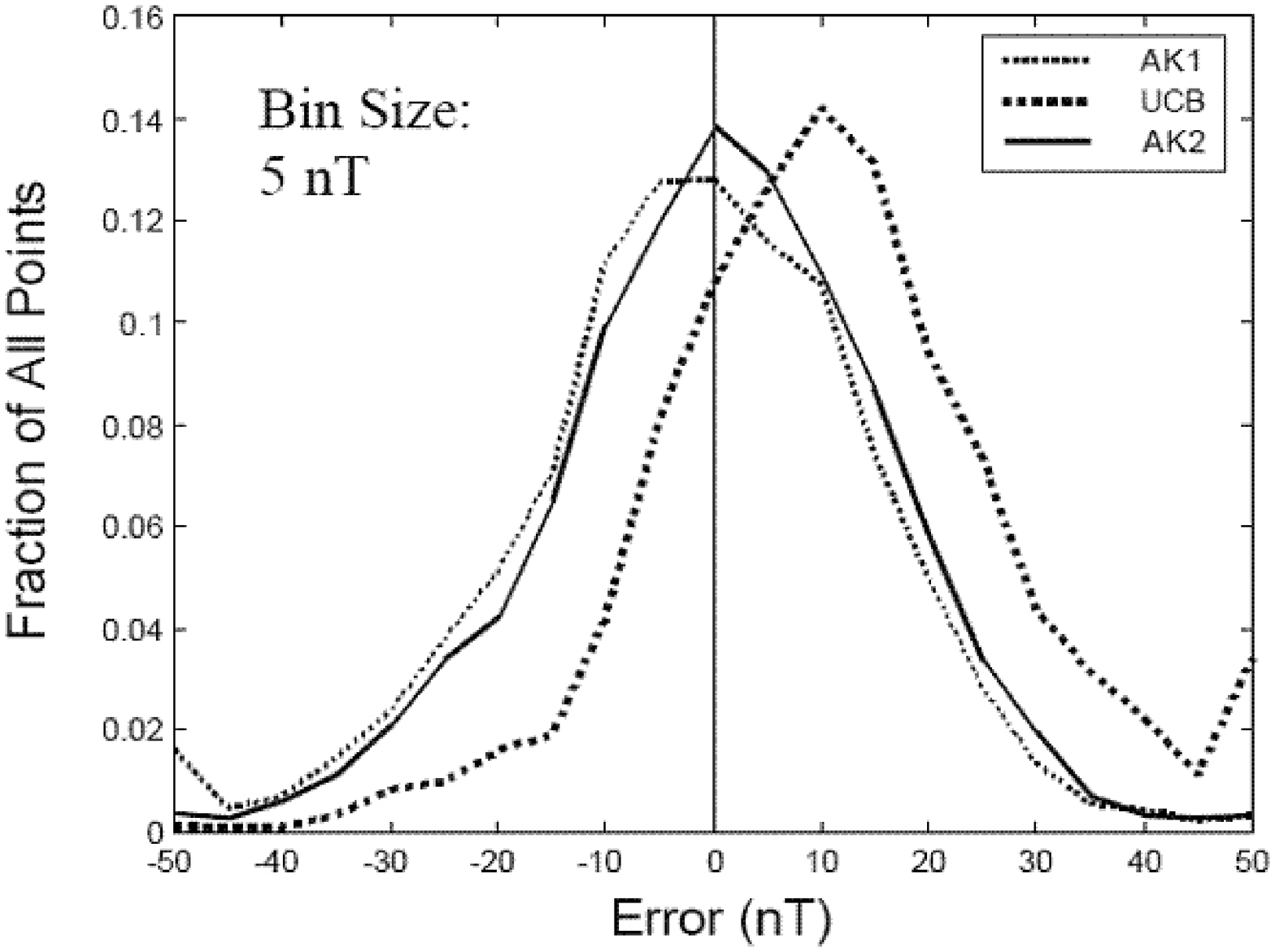}
\includegraphics[width=\columnwidth]{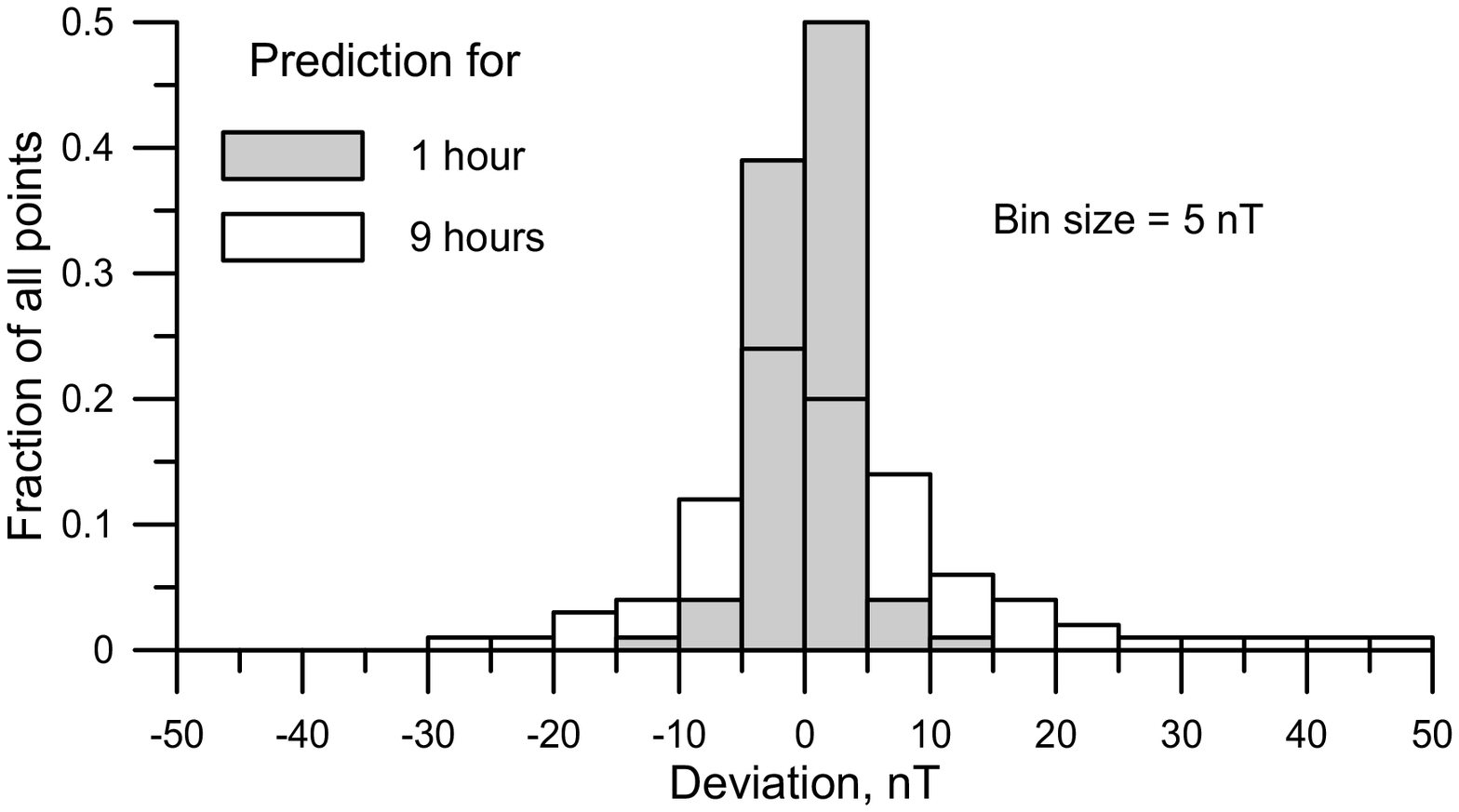}
\caption{Error charts of predictions results of \cite{ref:OM1} (top) and ours (bottom) 1 hour ahead. Error chart for our 9 hour prediction is plotted for reference}
\label{fig:16}
\end{figure}

Now let us discuss parameters, whose geoeffectiveness was determined by this
method, and demonstrate that they are indeed geoeffective.

Latitudinal flow angle $\theta_V$ was mostly associated with the southern
component of IMF. I plotted the distribution of its value and the corresponding
mean Dst value on Fig. \ref{fig:9}. The distribution looks similar to a normal
distribution, but it significantly differs from the normal one according to
$\chi^2$ test. This manifests in much larger number of points with deviations
more than $\sigma$ than follows from the normal distribution. This is mostly
caused by the number of points in the wing bins
$\left\vert\theta_V-\left<\theta_V\right>\right\vert>4\sigma$ being equal to
196 points versus 11 points in the case of normal distribution. However, most
of these points were obtained in the 1960s, when quality of measurements was
much worse then today. This period includes the maximum and minimum values of 
$\theta_V$, equal to $-59.7^\circ$ and $18.8^\circ$. Nevertheless, these
points constitute only a minor fraction of all points and didn't affect the
linear regression routine. Assuming normal distribution we obtain
$\sigma=2.925$ and $\left< \theta_V \right> =0.27<0.1\sigma$. Thus, the
distribution is insignificantly shifted towards positive values.

\begin{figure}[tb]
\includegraphics[width=\columnwidth]{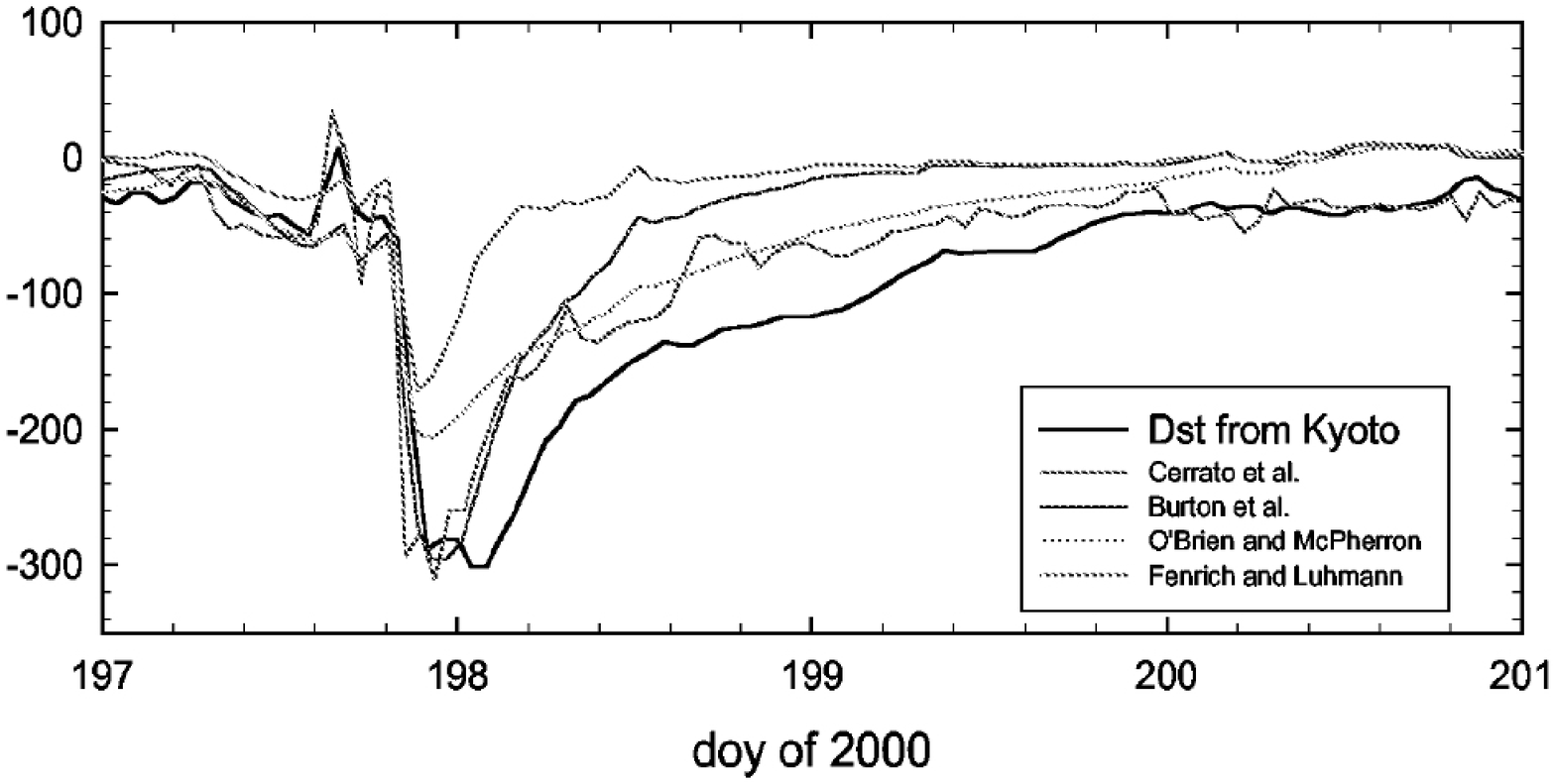}
\includegraphics[width=\columnwidth]{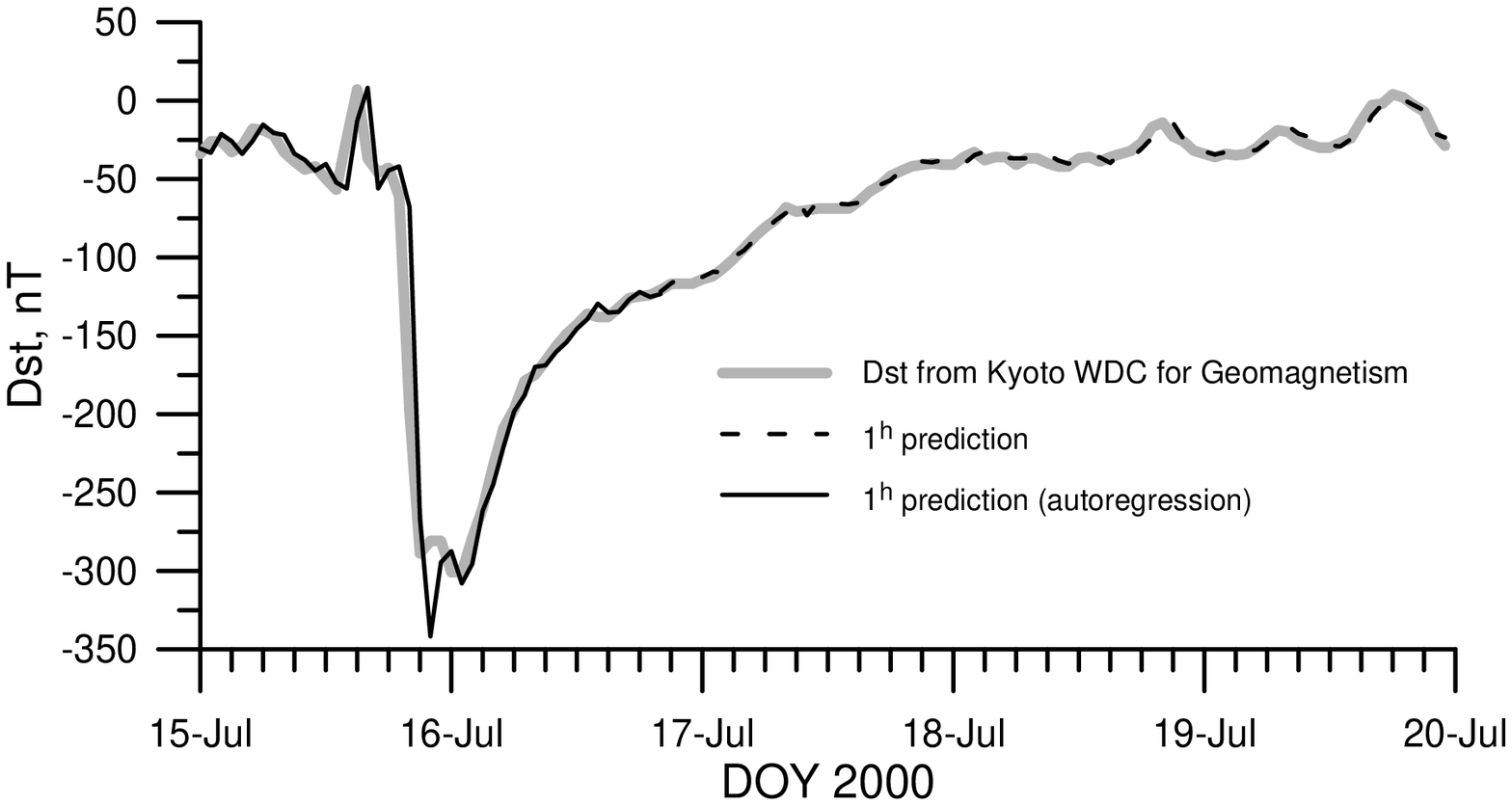}
\includegraphics[width=\columnwidth]{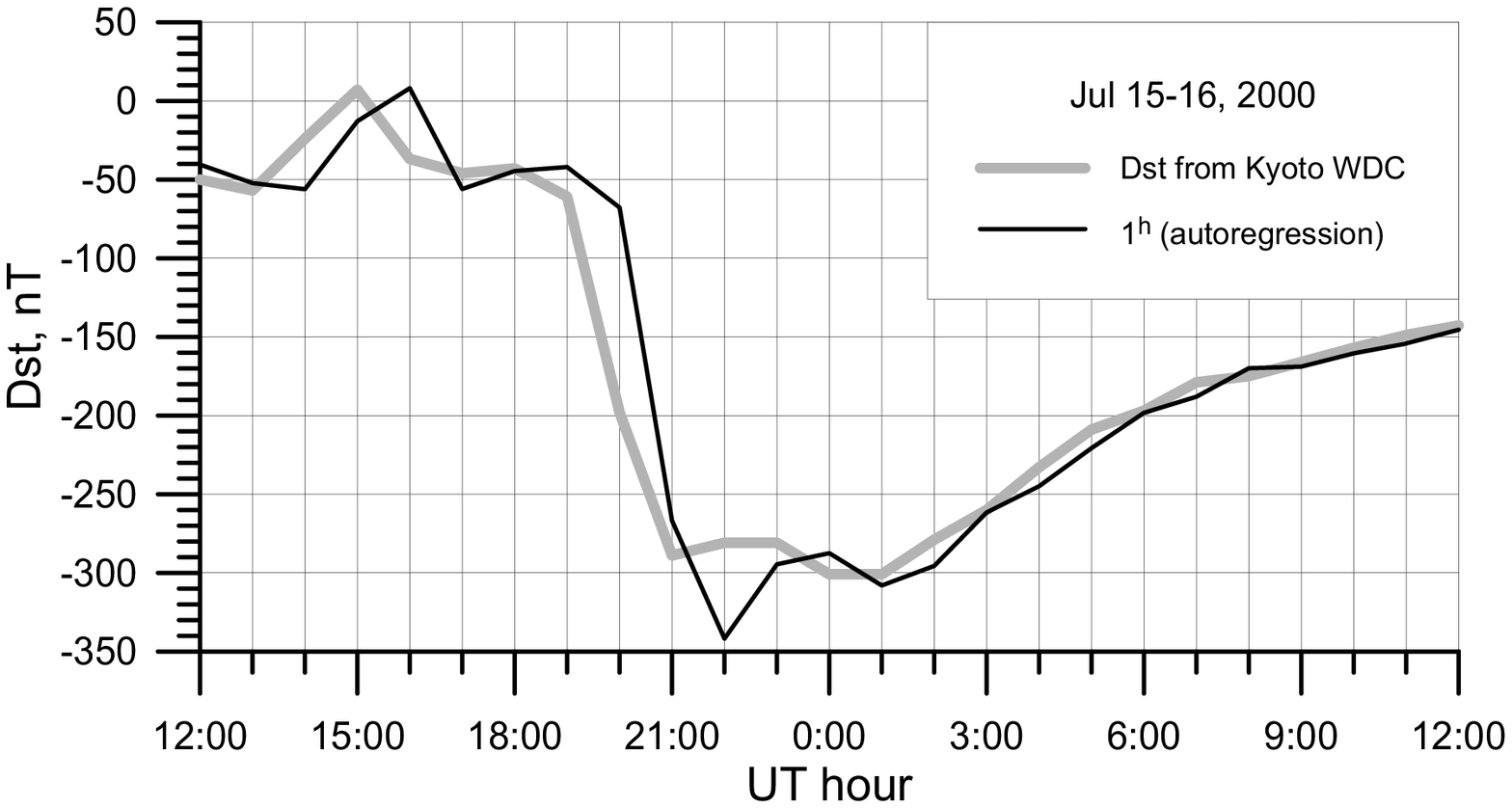}
\caption{Comparison between predictions results of \cite{ref:Cerrato}, \cite{ref:BMR}, \cite{ref:OM1}, \cite{ref:FL} (top) and ours (middle) 1 hour ahead. Bottom plot is a scaled up version of the middle one}
\label{fig:17}
\end{figure}

\begin{figure}[tb]
\includegraphics[width=\columnwidth]{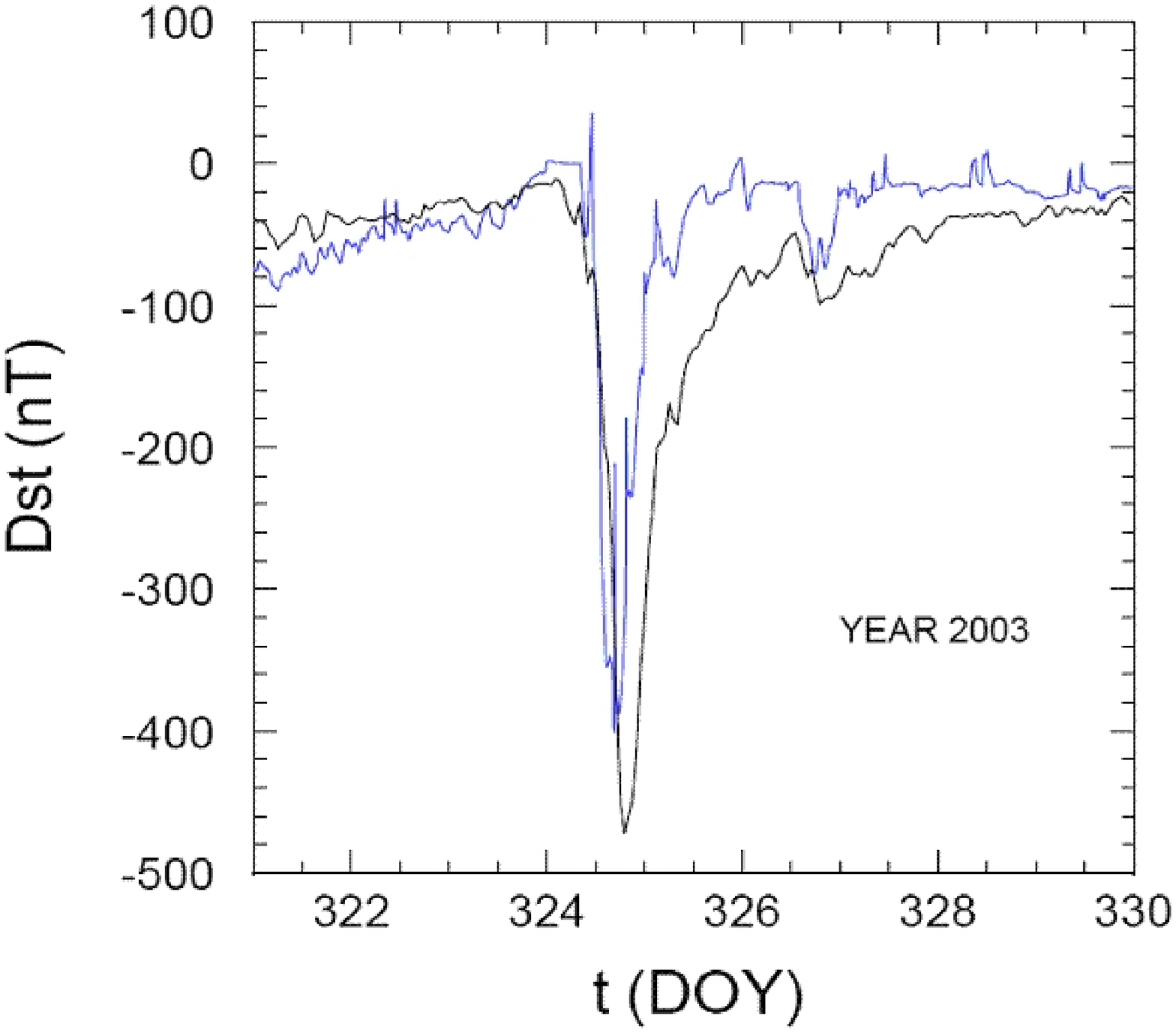}
\includegraphics[width=\columnwidth]{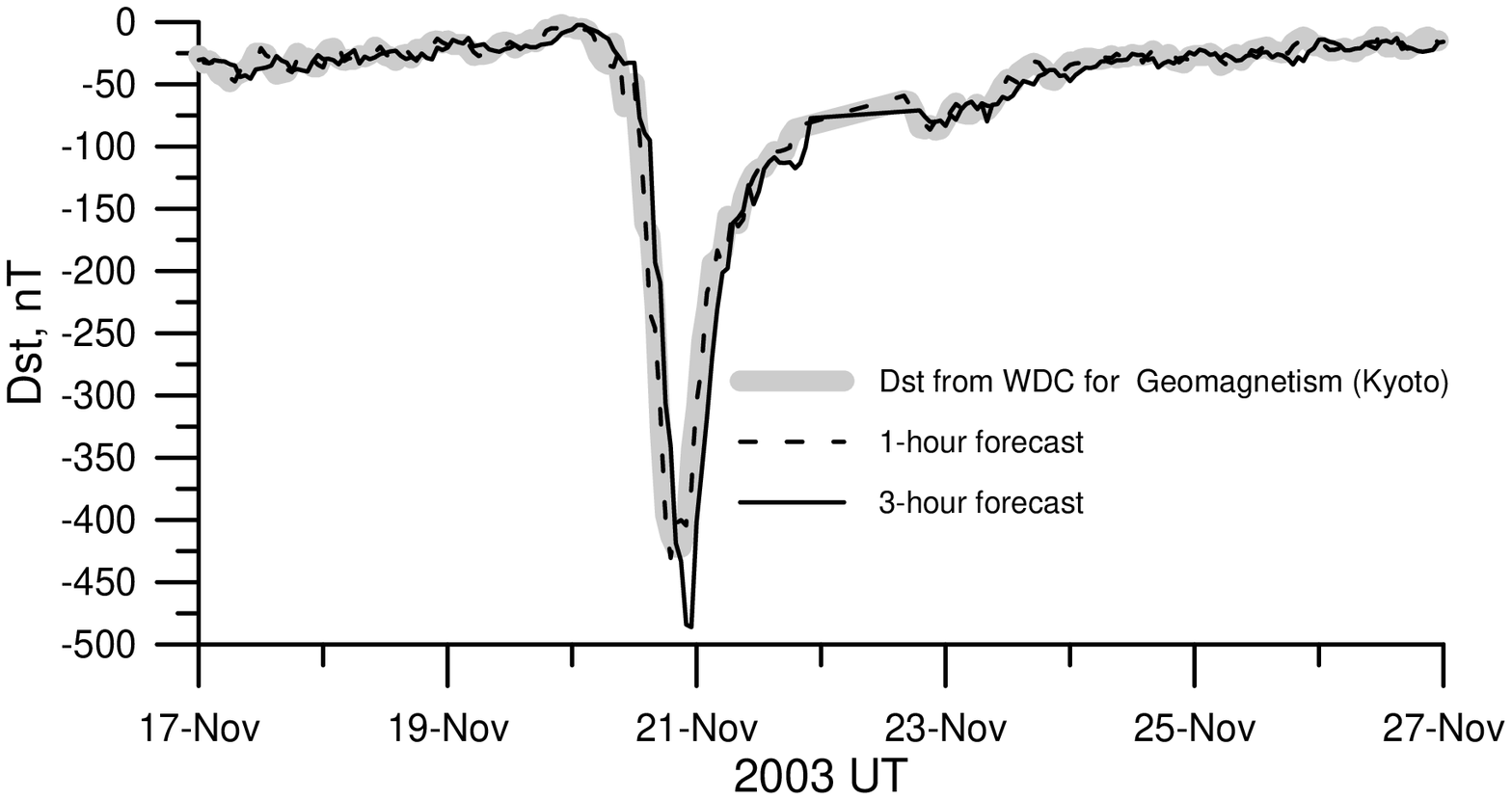}
\caption{Comparison between predictions results of \cite{ref:Pallocchia} (top) and ours (bottom) 1 hour ahead. Our 3-hour prediction is given for reference. On the top plot black line depicts Dst from Kyoto WDC, and the blue line -- 1 hour prediction}
\label{fig:18}
\end{figure}

If we ignore the wing bins in the distribution of mean Dst values against
$\theta_V$, which are somewhat random due to small amount of points in them, we
will notice a slight almost linear trend. If we plot the sum of terms
containing $\theta_V$ (Fig. \ref{fig:10}), we will notice a similar trend. If
we select two subsamples, one $-8<\theta_V<-5$ and other $4<\theta_V<9$, and
verify the hypothesis that the difference between the corresponding average Dst
values is statistically significant using a one-sided Student test, we obtain 
$t_\infty=6.278$, which is over 99.95\% significant.

If we divide the sample in two subsamples: one for northern summer and one for
northern winter, we obtain such a picture (Fig. \ref{fig:11}). Once again, let
us not look at the wing bins. What do we see? Summer distribution has an
obvious linear trend, but the winter one has not. If we apply the Student test
to the same intervals now, we obtain that $t_\infty$ is 5.44 in the summer and
only 0.059 in the winter. The prior corresponds to more than 99.95\%
significance, while the latter -- to less than 10\%. This could mean that there
are two factors connected with the latitudinal flow angle, which work together
in the summer and against each other in the winter. The physical explanation of
this phenomenon, however, lies beyond the scope of this paper.

Longitudinal flow angle $\varphi_V$ was only occasionally used in models.
However, it appeared to be even more significant than the latitudinal flow
angle. Its distribution together with corresponding mean Dst values is plotted
on Fig. \ref{fig:12}, where white bars show the complete dataset sans rejects,
and the grayed bars show the quiet-time sample with $Dst>-50 nT$. Like the
latitudinal flow angle, the distribution of the longitudinal flow angle
resembles normal distribution. However, $\chi^2$ test disproves the relevant 
null-hypothesis. Once again, this is mostly due to wing bins which are mostly
formed of data points, corresponding to measurements in 1960s, including
maximum and minimum values equal to $-65.6^\circ$ and $48.5^\circ$. Assuming
normal distribution we obtain $\sigma=2.934$ and
$\left<\varphi_V\right> =-0.30\approx -0.1\sigma$.

A significant trend is the most prominent feature of this figure. If we plot a
sum of regressors, which contain $\varphi_V$ (Fig. \ref{fig:13}), we see a very
similar trend. Like before, we plot the distribution for summer and winter
subsamples separately (Fig. \ref{fig:14}). We see that the trend is identical
on both plots, so the corresponding effect is season-independent. The list of
regressors for $k=1$, containing $\theta_V$ and $\varphi_V$, is given in Table
\ref{tbl:1}. It contains the regressors themself, their coefficients and $F$
values.

Thus, we demonstrated that our method is truly capable of pointing out new
geoeffective parameters and verified the geoffectiveness of two such quantities.

\section{Prediction results}\label{s:Results}

\begin{table}
\caption{Regressors containing the flow angles. $V(j)$ is the bulk flow velocity of the solar wind}
\label{tbl:1}
\begin{tabular}{rc@{~}cc}
\tableline
$i$&$x_i$&$C_i$&$F_i$\\
\tableline
$1$& $\theta_V(j)\cdot V(j)$&                 $\phantom{-}(2.8\pm 0.9)\cdot 10^{-5}$&          $10.2$\\
$2$& $\theta_V^4(j)$&                                   $(-4.9\pm 1.2)\cdot 10^{-5}$&          $17.5$\\
$3$& $a_1(j)\cdot \theta_V(j)\cdot D_{st}(j)$&$\phantom{-}(1.3\pm 0.3)\cdot 10^{-3}$&          $18.7$\\
$4$& $a_1(j)\cdot \theta_V^4(j)$&                       $(-3.2\pm 1.6)\cdot 10^{-5}$&$\phantom{0}4.0$\\
$5$& $\varphi_V(j)$&                                    $(-4.1\pm 0.6)\cdot 10^{-2}$&          $50.1$\\
$6$& $\varphi_V^2(j)$&                                  $(-4.5\pm 1.5)\cdot 10^{-3}$&$\phantom{0}9.3$\\
$7$& $\varphi_V^3(j)$&                                  $(-3.5\pm 1.3)\cdot 10^{-4}$&$\phantom{0}7.1$\\
$8$& $\varphi_V^4(j)$&                        $\phantom{-}(5.3\pm 1.7)\cdot 10^{-5}$&          $10.1$\\
$9$& $a_2(j)\cdot \varphi_V(j)$&                        $(-2.2\pm 0.8)\cdot 10^{-2}$&$\phantom{0}7.9$\\
$10$&$a_2(j)\cdot \varphi_V^3(j)$&            $\phantom{-}(5.1\pm 1.9)\cdot 10^{-4}$&$\phantom{0}7.6$\\
$11$&$a_2(j)\cdot \varphi_V^4(j)$&                      $(-3.5\pm 1.4)\cdot 10^{-5}$&$\phantom{0}6.6$\\
\tableline
\end{tabular}
\end{table}

\begin{table}
\caption{Statistical characteristics of forecasting models}
\label{tbl:2}
\begin{tabular}{rrccc}
\tableline
$k$, h&RMS, nT&LC&PE&Note\\
\tableline
$1$& $3.76$& $\vphantom{\sum\limits_{}^{}{}}\frac{0.987}{0.977}$&$0.975$&\\
$1$& $4.50$& $\vphantom{\sum\limits_{}{}}\frac{0.982}{0.977}$&   $0.964$&autoregression\\
$1$& $3.15$& $\vphantom{\sum\limits_{}{}}\frac{0.977}{0.958}$&   $0.983$&$Dst>-50 nT$\\
$1$& $6.25$& $\vphantom{\sum\limits_{}{}}\frac{0.984}{0.963}$&   $0.931$&$Dst\le -50 nT$\\
$3$& $7.60$& $\vphantom{\sum\limits_{}{}}\frac{0.941}{0.906}$&   $0.899$&\\
$6$& $10.45$&$\vphantom{\sum\limits_{}{}}\frac{0.882}{0.813}$&   $0.809$&\\
$9$& $12.84$&$\vphantom{\sum\limits_{}{}}\frac{0.820}{0.727}$&   $0.711$&\\
$12$&$14.47$&$\vphantom{\sum\limits_{}{}}\frac{0.764}{0.662}$&   $0.636$&for reference\\
$18$&$16.72$&$\vphantom{\sum\limits_{}{}}\frac{0.677}{0.554}$&   $0.514$&for reference\\
$24$&$18.22$&$\vphantom{\sum\limits_{}{}}\frac{0.605}{0.505}$&   $0.423$&for reference\\
\tableline
\end{tabular}
\end{table}

Taking into account the considered parameters together with parameters, whose
geoeffectiveness was beyond doubts, like previous values of Dst, dawn-dusk
electric field, ram pressure of the solar wind and most of other parameters
from the OMNI2 database, we constructed models for predicting Dst 1, 3, 6, 9,
12, 18, and 24 hours ahead, and 3 more models for predicting Dst 1 hour ahead
for quiet and perturbed conditions and for the case when satellite data are
unavailable (autoregression, see more in \cite{ref:EPS08}). The statistical
characteristics of these models are summarized in Table \ref{tbl:2}. They
include Residual Mean Square (RMS), Linear Correlation Coefficient (LC), and
Prediction Efficiency ($PE=1-RMS^2/SD^2$, where SD is the sample's Standard 
Deviation). In divided cells the top number corresponds to the actual model,
and the bottom one -- to the simplest possible model $D_{st}(j+k)=D_{st}(j)$.
It is noteworthy that despites good correlation for all the models, in reality
only the 1-hour and 3-hour models are ready for practical use, and the 6-hour
model can potentially reach this state. This is due to a significant time shift
being present in further predicting models.

Note that since the proposed method is statistical, there is little difference
whether the ``training'' sample contains the period when prediction is made
(``test'' sample) or not. To further illustrate this point let us consider an
example: I predicted Dst 3 hours ahead using the ``test'' sample from 2007 to
2008 and 3 ``training'' samples: the sample from 1976 to 2003 gives LC=0.851,
the sample from 1976 to 2006 gives LC=0.854, and the sample from 1976 to 2008
gives LC=0.854 as well. The first two ``training'' samples do not contain the
``test'' sample, but the results are slightly different. The third ``training''
sample contains the ``test''sample, yet the result is the same as for the
second sample. This yields a conclusion that the volume and statistical
properties of the ``training'' sample affect the prediction results stronger
than the inclusion of the ``test'' sample. So, the inclusion of the ``test''
sample to the ``training'' sample has little or no effect on the LC. PE is
calculated independently from the ``test'' sample and is not affected by its
selection in any way.

\begin{figure*}[tb]
\includegraphics[width=\columnwidth]{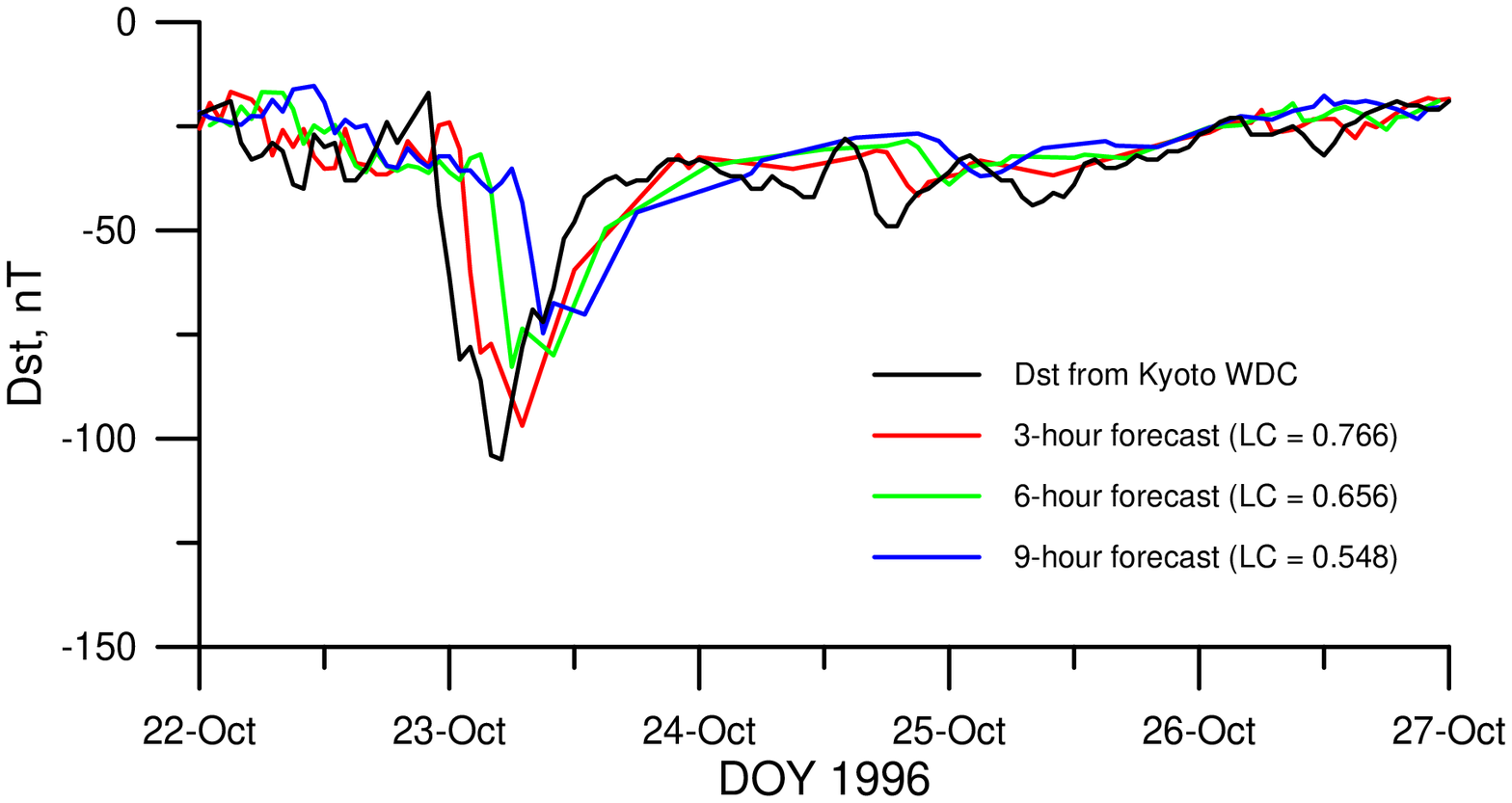}
\includegraphics[width=\columnwidth]{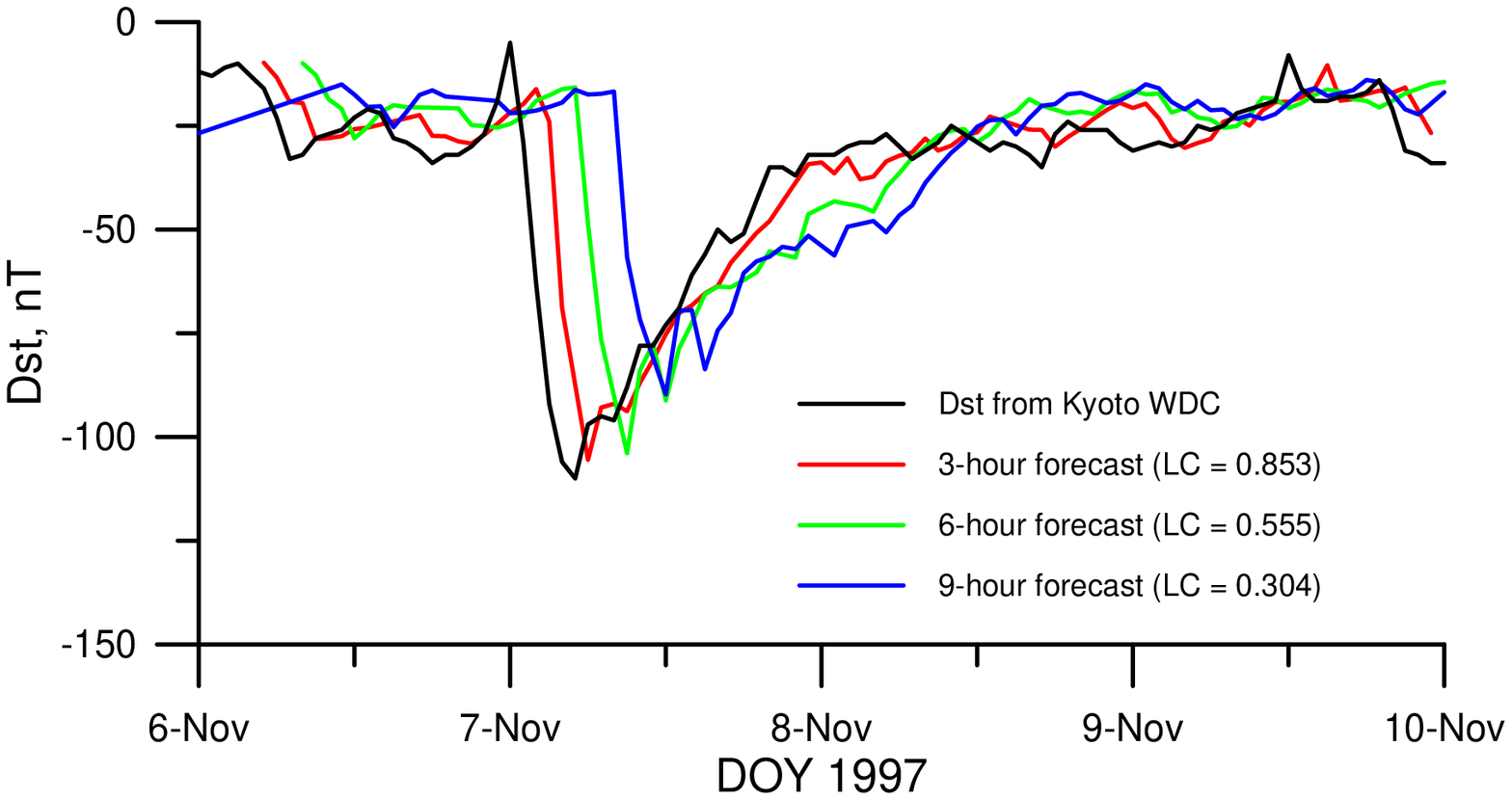}
\includegraphics[width=\columnwidth]{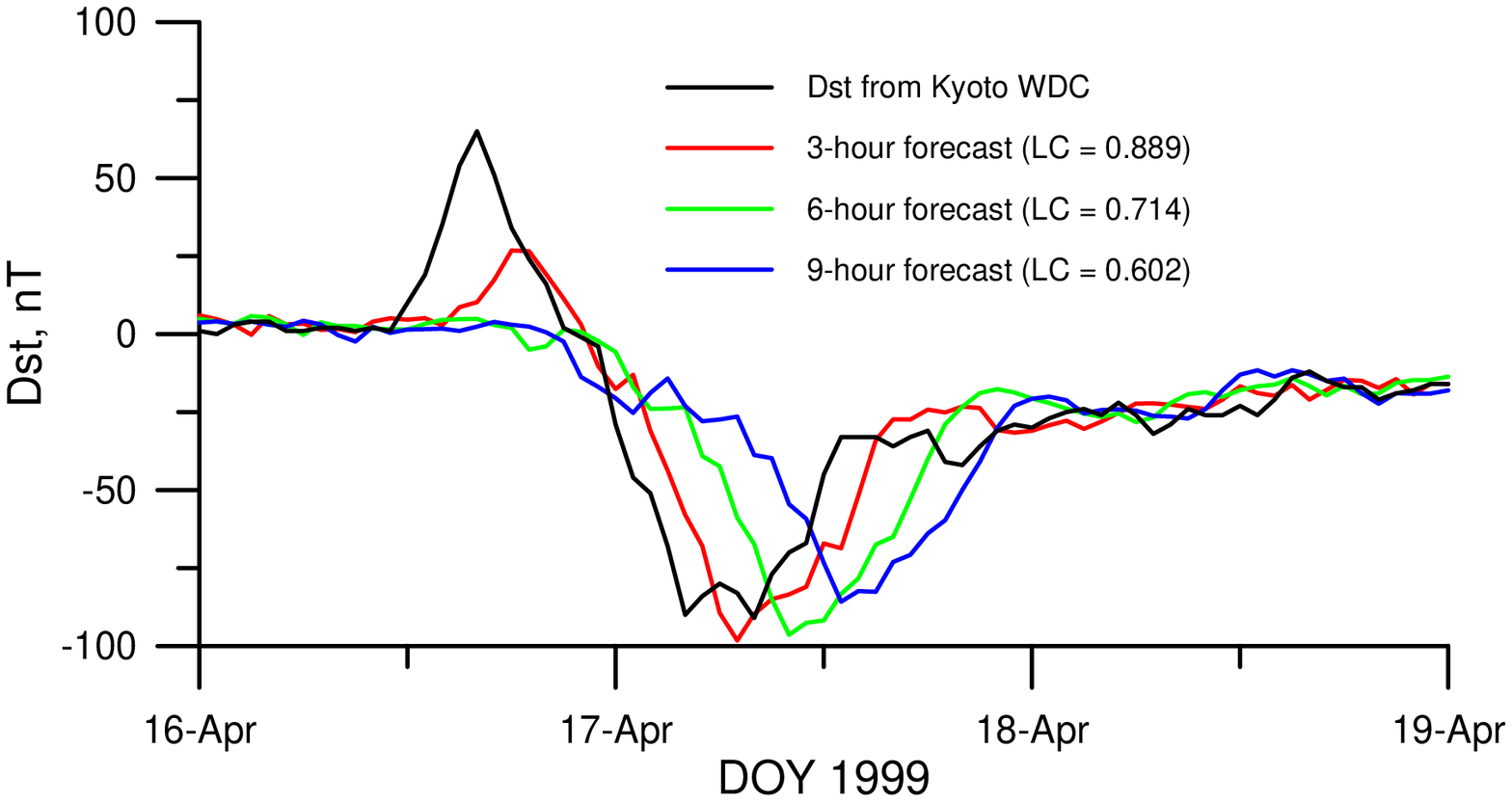}
\includegraphics[width=\columnwidth]{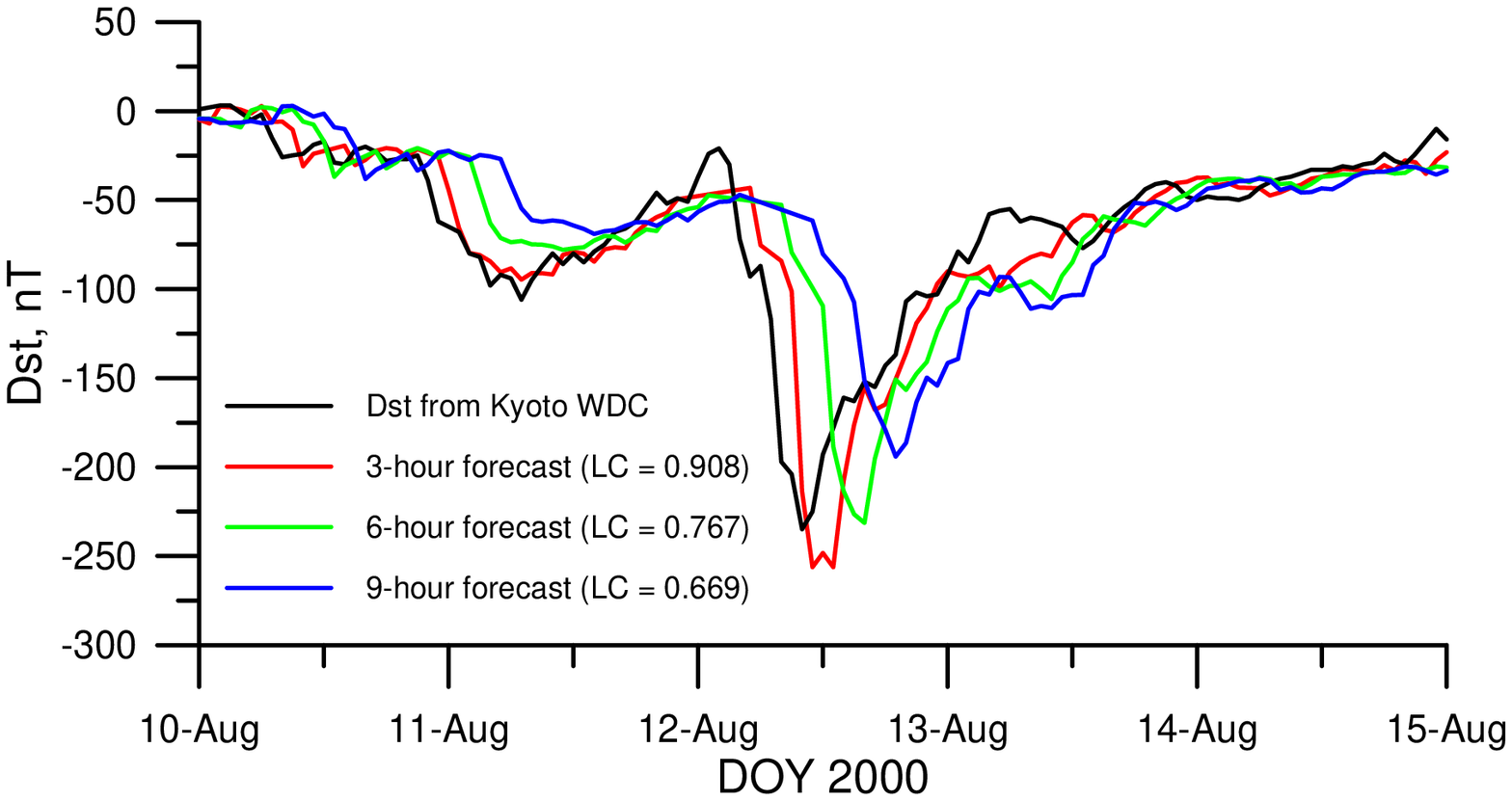}
\includegraphics[width=\columnwidth]{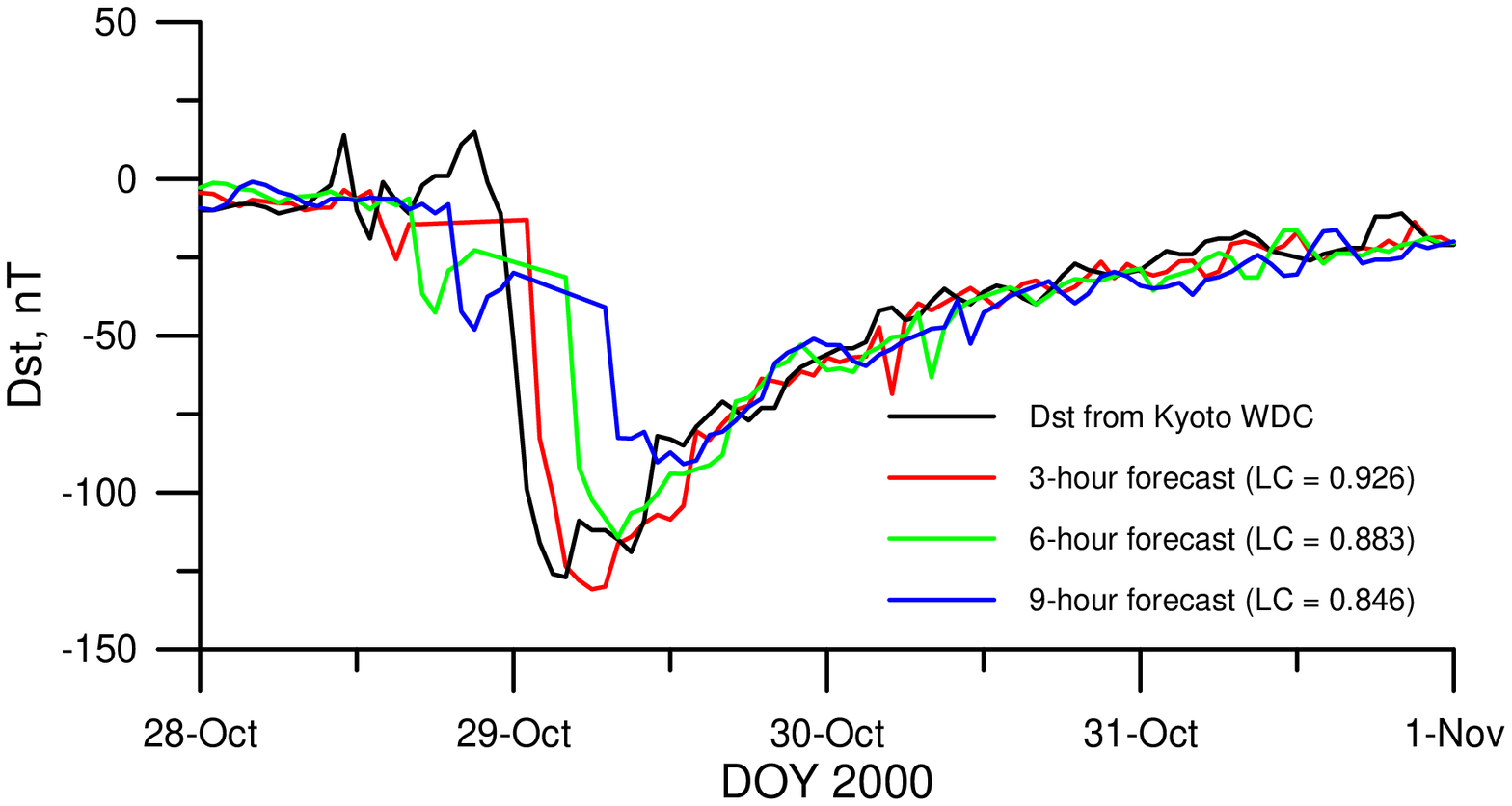}
\includegraphics[width=\columnwidth]{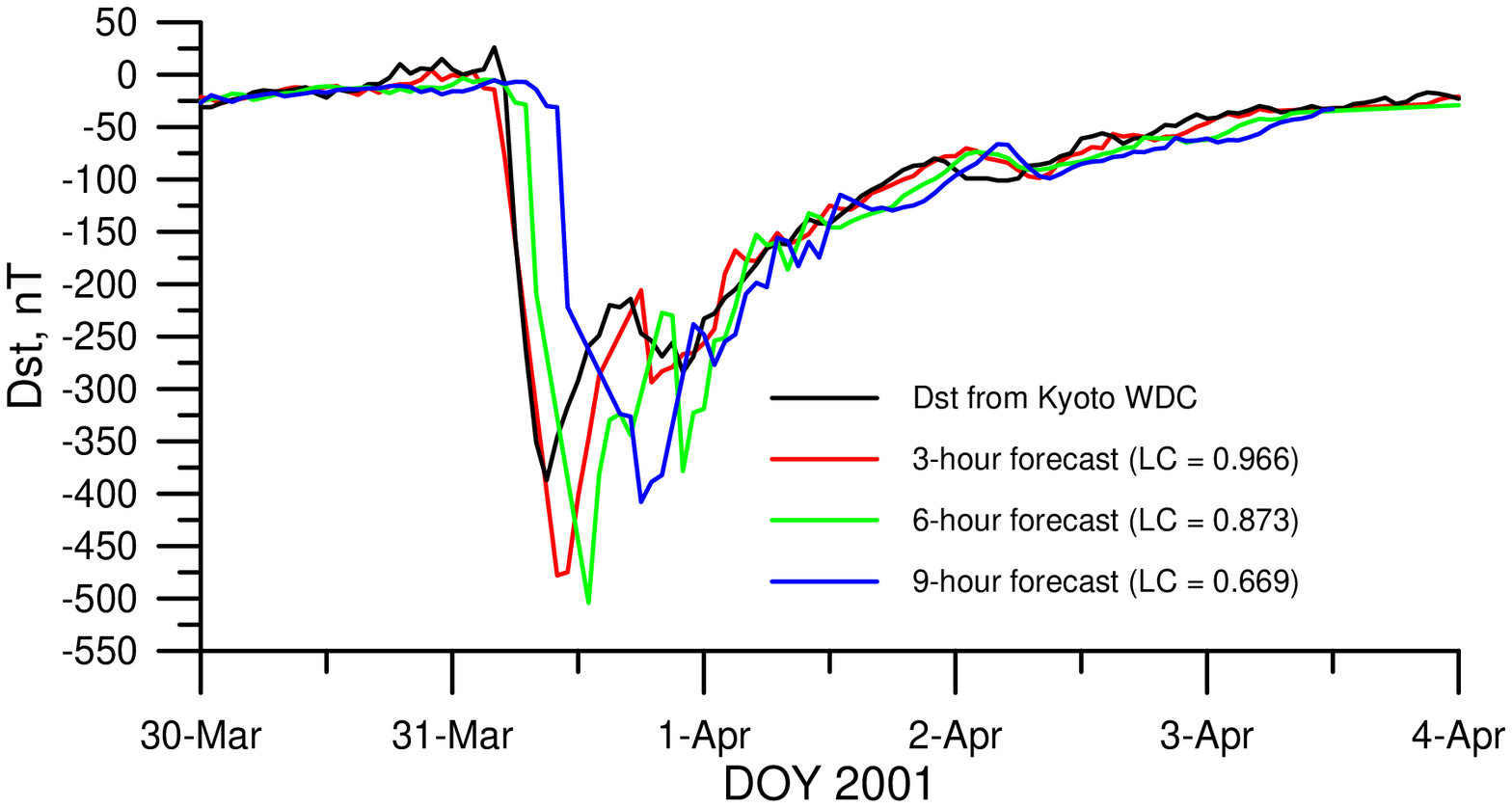}
\phantom{1234567890}
\caption{Prediction results for some specific intervals}
\label{fig:19}
\end{figure*}

We also present graphical representations of prediction in comparison with
results of other authors: \cite{ref:OM1} (Fig. \ref{fig:15}, \ref{fig:16}),
\cite{ref:Cid} (Fig. \ref{fig:17}), and \cite{ref:Pallocchia} (Fig.
\ref{fig:18}). Note, that the intervals for prediction were selected by authors
of the original papers, and the coefficients in eq. (\ref{eqn:1}) were the same
all the time and for all the figures. It is clearly visible that our method
provides much more precise forecast than most empirical models and typical
neural network models. Ridiculously, even our $9^h$ model is more precise than
most empirical $1^h$ models. The autoregression model, described by eq. (\ref{eqn:2}),
though, lags in the left part of the plot due to a rapid positive change of Dst
at 1500 UT. The lag persists through the growth phase and the main phase, and
vanishes only in the recovery phase. For this reason, the autocorrelation model
holds little practical value and should be considered as a transitional result,
required to construct the full model. It is, however, possible to improve it by
adding terms describing temporal variations, and, for example, the number of
sunspots, but then the term ``autoregression'' will no longer be applicable.

On Fig. \ref{fig:19} we present the results of prediction 3, 6 and 9 hours
ahead for a number of events, kindly selected for us by V.G. Fainshtein,
which are particularly hard to predict by medium-term methods, such as \cite{ref:EFR},
to verify the efficiency of our method. We can see that this method's accuracy
is higher for stronger storms, which are of greater interest. A huge advantage
of this method is that the most resource-demanding operation -- the calculation
of the regression coefficients -- should be performed only once for each model.
The prediction itself is just a summation of a polynom, which usually takes no
more than 4-6 seconds on an average PC (including disk I/O), which allows for
creation of fully automated operational on-line space weather forecast services.

\section{Conclusion}\label{s:Conclusion}

The proposed regression approach appeared to be more than adequate for space 
weather forecasting. For the forecasting per se, its main advantages are quite
good correlation (about 90\% for 6 hours forecast), adaptability to any samples,
and very fast forecasting code (typically about 5 seconds on an average PC).
For the identification of geoeffective parameters it is extremely convenient
and easy to use. In particular, it allowed to uncover 2 new geoeffective
parameters -- the latitudinal and the longitudinal flow angles of the solar
wind.

This is just a short summary of the regression modeling method, since its full
description would take much more space. Of course, this method can be used in
conjunction with other methods, first of all, with physical methods of
detection of large-scale perturbations in the solar wind and with empirical
models.

\acknowledgments
The author would like to thank Prof. O.K. Cheremnykh, Prof. V.A. Yatsenko,
and Academician V.M. Kuntsevich for fruitful discussion, Prof. V.G. Fainshtein
for useful remarks and for providing a list of geomagnetic events for
validation of the model, and Reviewer \#1 for valuable comments which greatly
improved this article.

The author is grateful to the Space Physics Data Facility (SPDF) and the 
National Space Science Data Center (NSSDC) for the free online OMNI2 
catalogue and to the World Data Center for Geomagnetism (WDC-B) at Kyoto 
University for the free online catalogue of geomagnetic indices.

This work was partially supported by the National Academy of Sciences of 
Ukraine, state programme ``GEO-UA'', and by the National Space Agency of 
Ukraine, state contract No. 8-09/08 ``Programa-N''.

\end{document}